\documentclass[12pt,a4paper,twoside,openright]{revtex4}
\usepackage[english]{babel}
\usepackage{array,booktabs}   % http://ctan.org/pkg/{array,booktabs}
\usepackage{array} 
\usepackage{amsmath} 
\usepackage{relsize}
\usepackage{wrapfig}
\usepackage{calc}
\usepackage{sitem}
\usepackage{pdflscape}
\usepackage{color}
\usepackage{float}
\usepackage{graphicx}
\usepackage[nodisplayskipstretch]{setspace}

\usepackage{setspace}
\usepackage{tabularx}

\usepackage{color}
\usepackage{epsfig}
\usepackage{epstopdf}
\usepackage{amssymb}
\usepackage{epstopdf}
\usepackage{appendix}
\usepackage{romannum}
\usepackage{soul}

\definecolor{blizzardblue}{rgb}{0.67, 0.9, 0.93}
\definecolor{bubblegum}{rgb}{0.99, 0.76, 0.8}
\usepackage[urlcolor=blizzardblue]{hyperref}
\hypersetup{
	colorlinks=false,
	linkcolor=blue,
	filecolor=magenta,      
	urlcolor=bubblegum,
}
\begin{document}
	\pagenumbering{arabic}
	\title{Uncertainties due to hadronic production in final-state interactions at  long-baseline neutrino facility  }
	\author{ Ritu Devi$^{1}$\footnote{E-mail: rituhans4028@gmail.com}, Jaydip Singh$^{2}$\footnote{E-mail: jaydip.singh@gmail.com}, Baba Potukuchi$^{1}$\footnote{E-mail: baba.potukuchi@gmail.com}}
	
	\affiliation{Department Of Physics, University of Jammu, Jammu, India$^{1}$}
	\affiliation{Department Of Physics, University of Lucknow, Lucknow, India.$^{2}$}

	\bigskip
	\begin{abstract}

		Recent neutrino oscillation experiments used high atomic number nuclear targets to attain sufficient interaction rates. The use of these complex targets introduced systematic uncertainties due to the nuclear effects in the experimental observables and need to be measured properly to pin down the discovery. Through this simulation work, we are trying to quantify the nuclear effects in the argon (Ar) target in comparison to hydrogen (H)  target which are proposed to be used at  Deep Underground Neutrino Experiment far detector  and near detector, respectively.  Generates Events for Neutrino Interaction Experiments and NuWro, two neutrino event generators, are used to construct final state kinematics.  To quantify the systematic uncertainties in the observables, we present the ratio of the oscillation probability (P(Ar)/P(H)) as a function of reconstructed neutrino energy. \\
	\textit{Keywords:} {Neutrino, uncertainties, final-state interactions, survival probability, nuclear effects.}

	\end{abstract}

	\maketitle
	\section{Introduction}
	
	The proposed  Deep Underground Neutrino Experiment (DUNE)  is a long-baseline neutrino oscillation experiment \cite{1,1DUNE,2DUNE,3DUNE}. The primary scientific goal of this experiment is to make precise measurements of the parameters governing neutrino oscillations and constraining the CP violation phase in the leptonic sector \cite{4DUNE}. Another scientific goal of the DUNE project is the determination of mass hierarchy, either it is  normal hierarchy (NH) or inverted hierarchy (IH) which is not known yet from experiments of supernova physics \cite{1b}, atmospheric neutrino physics, and proton decay physics \cite{1a}. DUNE will have neutrino and anti-neutrino spectrum ($ \nu_{\mu}(\bar{\nu_{\mu}}), \nu_{e}(\bar{\nu_{e}}),$ and $\nu_{\tau}(\bar{\nu_{\tau}})$) beam facility at Fermilab that will be characterized by a near detector (ND) near the neutrino source at Fermilab and also by far detector(FD) at Sanford Underground Research Facility (SURF) in South Dakota, USA. This will provide a baseline facility of length 1248 km to study the matter effect. Both the detectors will have Argon (Ar) target material to observe the neutrino spectrum that will help to overcome various systematic uncertainties. ND will be observing the un-oscillated neutrinos spectrum while the FD will be observing the oscillated neutrinos spectrum. Precise energy measurement of each flavor at both the detectors is crucial for achieving the scientific goal planned for the experiment. Oscillation probabilities as a function of reconstructed neutrino energy are measured by comparing the events rate as a function of reconstructed neutrino energy at  both ND and FD detectors, which are used to constrain values of the neutrino oscillation parameters and also proton lifetime measurement. These uncertainties cause major challenges in the determination of the CP-violating phase which is yet to be constrained. Precise CP phase value is not only crucial for verifying the baryon asymmetry of the universe but also for explaining physics beyond the standard model and effective mass of neutrinos \cite{jd_eff}.  
	
	Neutrinos are weakly interacting massive (extremely tiny mass) particles that interact very rarely with matter. Due to its weakly interacting nature, it is very hard to detect them in ordinary detectors. The detector thus  proposed by  DUNE collaboration is an extremely advanced detector of this generation compared to other baseline neutrino experiment detectors. This multipurpose DUNE  detector will have a modular structure and each module has different technology. Each detector will be capable of detecting neutrino events and other rare phenomena. Neutrino interacts with the charged current(CC) and neutral current (NC) processes. Interaction event with the CC process produces leptons ( $e^{+},e^{-},\mu^{+}, \mu^{-}, \tau^{+}, \tau^{-}$) corresponding to the neutrino flavors ($ \nu_{e}, \bar{\nu_{e}},\nu_{\mu}, \bar{\nu_{\mu}}, \nu_{\tau}, \bar{\nu_{\tau}}$) and hadrons in the final states. Interaction events with the NC produce the same neutrino flavor and hadrons in the final states. Those processes are not considered in our MC sample. Recent neutrino experiments are using a material with a high atomic number as detector materials in order to increase the interaction rates and that further adds systematic uncertainty in neutrino energy reconstruction due to nuclear effects. Major nuclear effects present in the nuclear targets are, uncertainties from the binding energy, multinuclear correlation, nuclear Fermi motion effects, and final-state interactions (FSI) of produced hadrons in different interaction channels. The requirement for accurate knowledge of neutrino oscillation parameters depends on many factors among which precise reconstruction of neutrino energy is of utmost importance. The neutrino oscillation probability itself relies on the energy of the neutrinos, and any inaccuracy of measurement of neutrino energy will be disseminated to the measurements of neutrino oscillation parameters, since it causes uncertainties in the cross-section measurement and event identification. These kinds of systematics can be constrained by using a simple targeted neutrino-hydrogen (H) scattering $(\nu(\bar{\nu})-H )$ data. 
	
	%check belo paragraph
	Besides, the uncertainties in the total neutrino cross-sections which adds  systematic uncertainties in the observables \cite{2_c}, the secondary particles that come out of the nucleus after the FSI also count up to systematic errors \cite{2_cc,2_d}. One of the main reasons is that the nucleons are neither at rest nor non-interacting inside the target nucleus. The intense neutrino beams in recent experiments resulted  greatly in an increase of statistics and decrease the statistical uncertainty. Now, these experiments appeal to the careful handling of systematic uncertainties. In this generation, neutrino experiments are using a complex nuclear target for interaction that result in sizeable and non-negligible nuclear effects. In most of the neutrino experiments, the nuclear effects are taken into consideration by Impulse Approximation \cite{5a} or Fermi Gas Model \cite{5b} when simulating neutrino scattering events. Recent experiments believe that nuclear effects need to be modeled carefully after looking at  experimental results of different neutrino experiments at different energy ranges.
	Neutrino experiments that used lower energy neutrino beams like MiniBooNE \cite{3d} and MicroBooNE \cite{2e} are sensitive to only two types of neutrino interactions i.e.  Quasi-Elastic (QE) and Resonance (RES). The pion production in these experiments makes up only 1/3 of the total neutrino-nucleon interaction cross-section. As we move towards the higher energy neutrino beam experiments such as NOvA, MINERvA, and DUNE, the pion production increases to two-third of the total cross-section \cite{5c}. Thus, it is important to measure the pion production channels carefully under good and quantitative control. One of the most important nuclear effects that gives rise to wrong events is due to the  FSI of hadrons which are produced in the reaction along with leptons \cite{jd}. The pions which are produced in neutrino-nucleon interactions while traveling inside the nucleus, can be absorbed, can change direction, energy, charge, or even produce more pions due to various processes like elastic and charge exchange scattering with the nucleons that exist in the nucleus through strong interactions. This leads to wrong  number of events depending on the emerging particles from the nucleus.

	The paper is organized in the following sections: the neutrino event generators GENIE and NuWro used in this work including a detailed comparison of the nuclear models incorporated in them are described in Section \ref{sec2}. The experimental details and simulations are described in Section \ref{sec3}, the  results, and the discussion are described in Section \ref{sec4}. Finally, our conclusions are  presented  in Section \ref{sec5}.

	\section{EVENT generators: GENIE and NuWro}
	
	\label{sec2}
	Event generators are used to generate the events for the detectors to perform the simulation-based results for any experiments to get an idea about the expected observables. Neutrino communities have their own event generators that have inbuilt models that are tuned based on experimental data and generators work as a bridge for experimental and theoretical physicists. We have selected the two most common neutrino event generators: Generates Events for Neutrino Interaction Experiments (GENIE)  v-3.0.6 \cite{3} and NuWro \cite{4a} version 19.01 to perform our studies that are currently being used by most of the neutrino experiments in the USA and other countries. GENIE  is   being  used  by  many  neutrino baseline experiments running around the world, such as   Minerva \cite{3b}, MINOS \cite{3c},  MicroBooNE \cite{3d}, NOvA \cite{3e}, and ArgoNEUT \cite{3g}  experiments. NuWro has been  developed by a group of physicists  Cezary  Juszczak et. al. at the Wroclaw University \cite{4a} and is now being used as a complementary neutrino events generator by most  neutrino experiments. GENIE has been developed putting special emphasis on scattering in the energy region of few GeV, important for ongoing and future oscillation studies experiments, while NuWro covers the neutrino energy scale from $\sim 100 $ MeV to $\sim 100 $ GeV. The basic architecture of NuWro follows better known MCs like NEUT \cite{4c} or GENIE \cite{3}. It operates all types of interactions important in neutrino-nucleus interactions as well as Deep Inelastic Scattering (DIS) hadronization and intranuclear cascade. NuWro serves as a tool to estimate the relevance of various theoretical  models being investigated now. GENIE is a modern and flexible platform for neutrino events simulation.
	
	In this section, we will go over the qualitative theoretical differences in nuclear models, as well as how neutrino-nucleus interaction processes are accounted for in both generators and some popular approaches to neutrino-nucleus interaction analysis. In GENIE, the Relativistic Fermi gas (RFG) model is used  which is based on the model suggested by A. Bodek and J.L. Ritche \cite{3h} while in NuWro, Local Fermi gas (LFG) is used as a nuclear model. Modeling of QE scattering is according to Llewellyn Smith model \cite{3i} in both the generators. Also the latest BBBA07 \cite{3n} and  BBBA05 \cite{3j} vector form factors are used by GENIE and NuWro, respectively. GENIE uses a variable value of axial mass $M_A$ between 0.99 - 1.2 $GeV/c^{2}$ while NuWro uses a  variable value of axial mass between 0.94 - 1.03 $GeV/c^2$. In our work, we have used axial mass $M_A$ = 1 $GeV/c^2$.  GENIE considered $\Delta$ contribution as well as other resonance modes individually based on Rein-Sehgal model \cite{3k} however, in NuWro we have  only $\Delta$ contribution and for the rest of resonance modes, we have an average based on Adler-Rarita-Schwinger model \cite{3o} for RES events. For DIS events, NuWro employs Quark-Parton model \cite{3p} while GENIE uses Bodek and Yang model \cite{3l}.
	
	\section{  Experimental details and simulations }
	\label{sec3}
	
	DUNE at Long Baseline Neutrino Facility (LBNF) will consist of ND and FD with same target material, but the difference is in dimensions and technology. The ND system will be located at Fermilab, 574 m  downstream and 60 m underground \cite{4DUNE} from the neutrino beam source point.

	The baseline design of the DUNE ND \cite{1} system comprises of three main components: (1) A 50t LArTPC called ArgonCube with  pixellated readout; (2) A multi-purpose tracker, the high-pressure gaseous argon TPC (HPgTPC), kept in a 0.5T magnetic field and surrounded by electromagnetic calorimeter (ECAL), together called the multi-purpose detector (MPD); (3) and an on-axis beam monitor called System for on-Axis Neutrino Detection (SAND). SAND \cite{1DUNE} will consist of  an  ECAL, a superconducting solenoid magnet, a thin active LAr target, a 3D scintillator tracker (3DST) as an active neutrino target, and a Low-density tracker to precisely measure particles escaping from the scintillator. The FD (and the similar ND) has to be made of heavy nuclei rather than hydrogen that is  the main  source of complication.  The primary goal of SAND is constraining systematics from nuclear effects.
	For a better understanding of nuclear effects in neutrino-nucleus interactions, it is useful to examine  what would happen if the detectors will be made of hydrogen because in a detector made of hydrogen, the initial state is a proton  at rest and there is no FSI. The 40 kt DUNE FD  will comprise  of four separate LArTPC detector modules, each with a fiducial volume (FV) of at least 10 kt, installed $\sim $1.5 km underground at the SURF \cite{D2}. The internal dimensions  15.1 m (w) $\times$ 14.0 m (h) $\times$ 62.0 m (l) of each LArTPC fits inside an FV containing a total LAr mass of about 17.5 kt. Detailed detector simulation and reconstruction analysis is done by the DUNE-collaboration that can be found in detail in references \cite{4DUNE, D3}. We have applied the same detector efficiency from figure 4.26 of Ref. \cite{D3}  and energy resolution 18\% for muon and 34\% for hadrons for the particles available in the FSI while reconstructing the neutrino energy in our analysis. 
	
\begin{figure}[hbt!]% figure* for wide figure, [h] [!] to change the placement
		\vskip1mm
\includegraphics[width=0.5\linewidth]{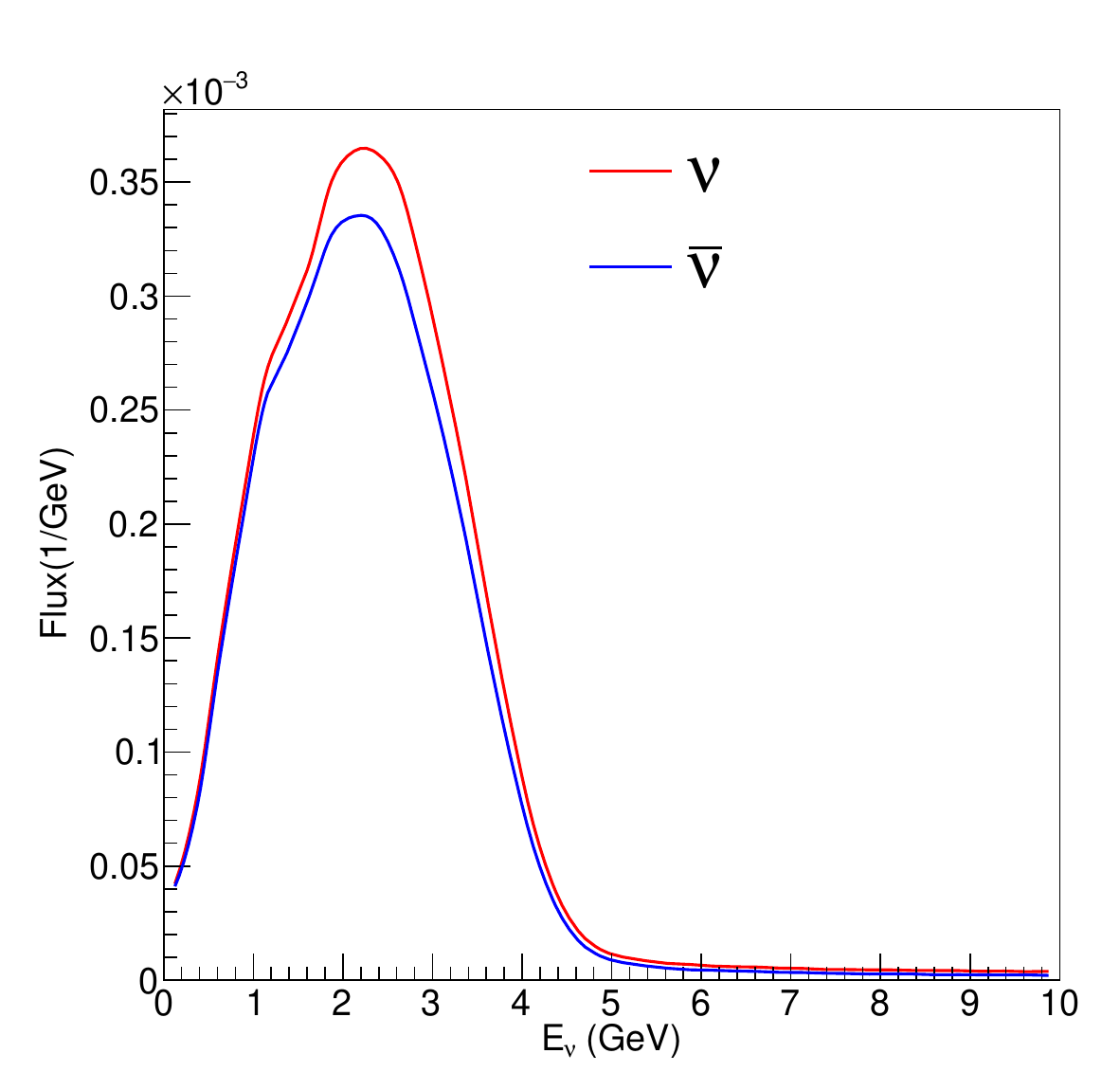} 
		\vskip-3mm\caption{The DUNE flux as a function of neutrino energy used in our work.  }
		\label{fig:1}
	\end{figure}

	We have used DUNE-ND  $\nu_{\mu}$ and $\bar{\nu_{\mu}}$ flux \cite{2_a}  having an energy range of 0.125-10 GeV shown in Figure \ref{fig:1} and  this flux is also folded with the oscillation probability to get the flux at the FD site. In this energy range, there is a mashing of several interaction processes viz. QE, RES, DIS, and two particle-two hole or Meson Exchange Current (2p2h/MEC). In order to do our analysis, we have generated an inclusive event sample of 1 million $ \nu_{\mu} (\bar{\nu_{\mu}})$-Ar,  and $\nu_{\mu}(\bar{\nu_{\mu}})$-H  using both GENIE and NuWro event generators. In our analysis, we took into account all of these interaction mechanisms.
	
	%Fig.~1

	\begin{figure*}[hbt!]% figure* for wide figure, [h] [!] to change the placement
		\vskip1mm
		\includegraphics[scale=0.4]{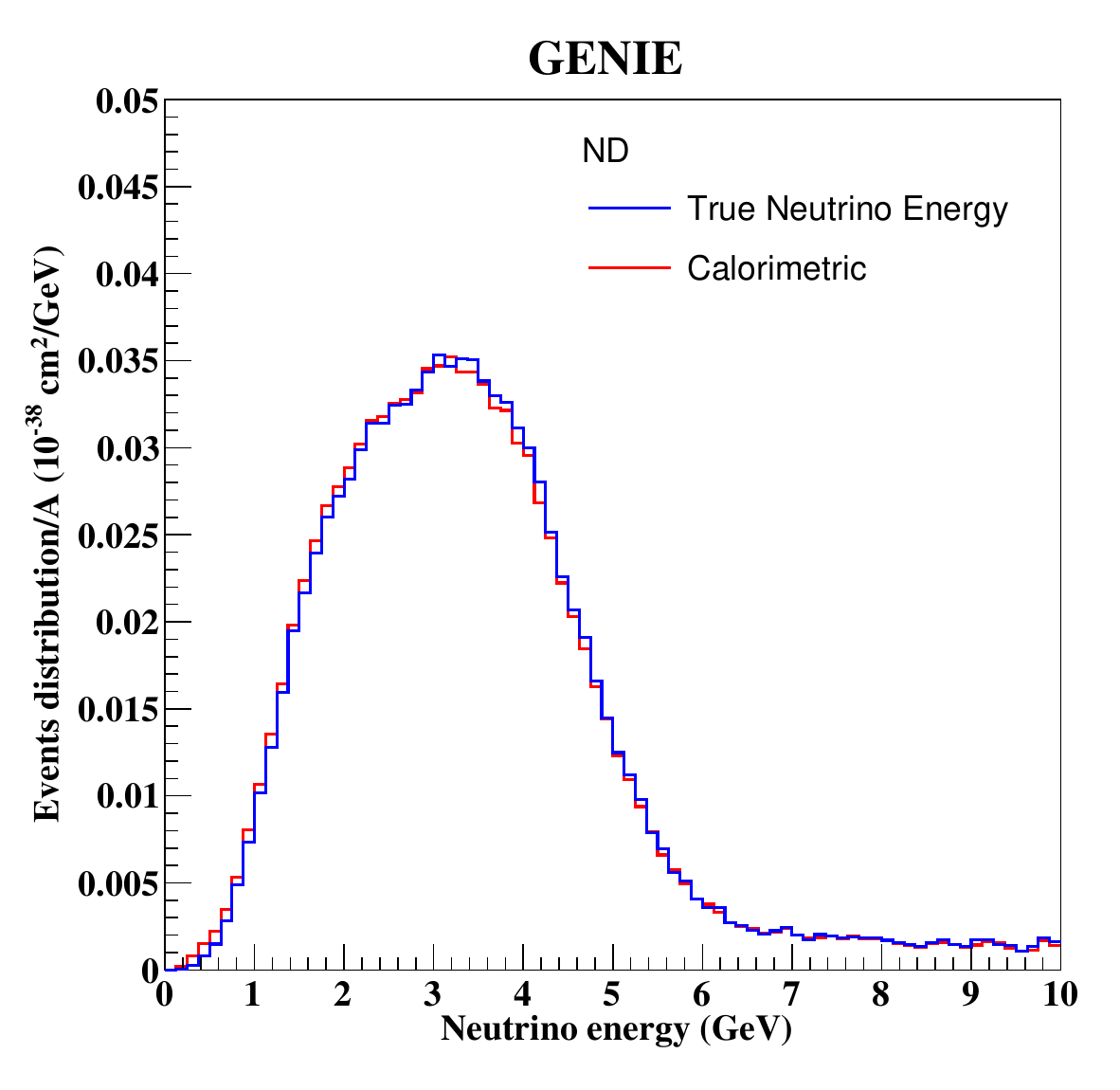}
		\includegraphics[scale=0.4]{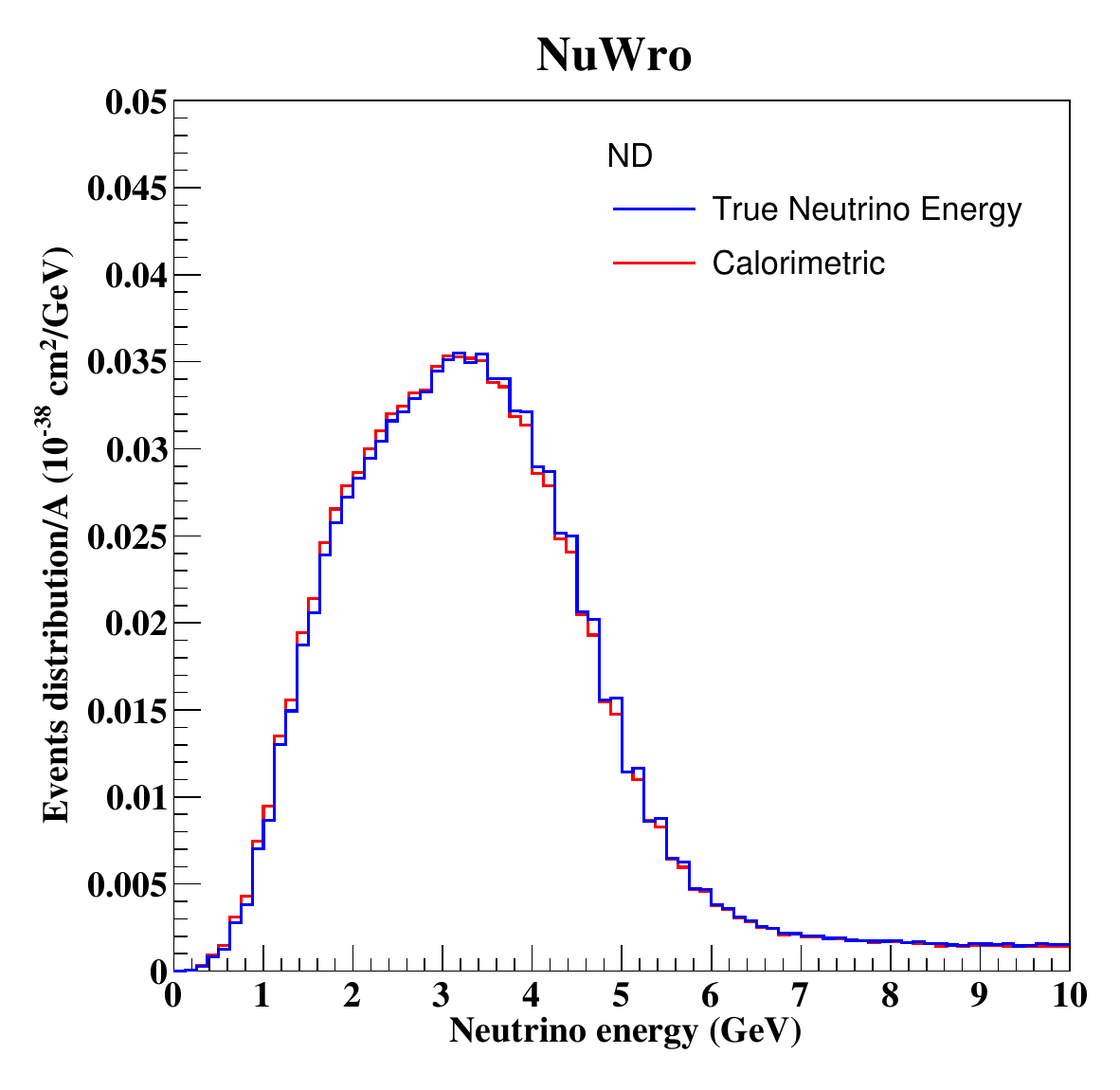}
		\includegraphics[scale=0.4]{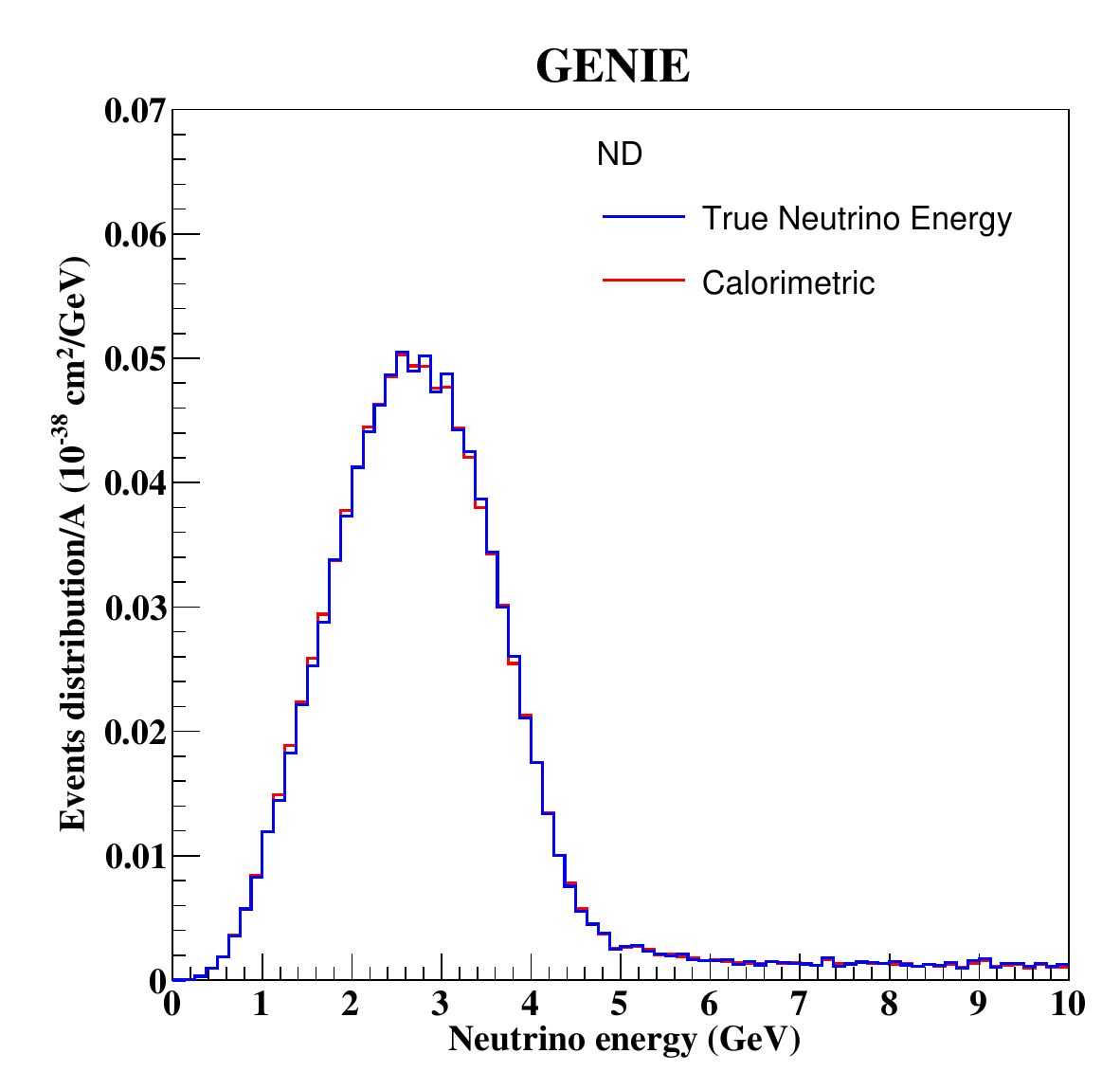} 	
		\includegraphics[scale=0.4]{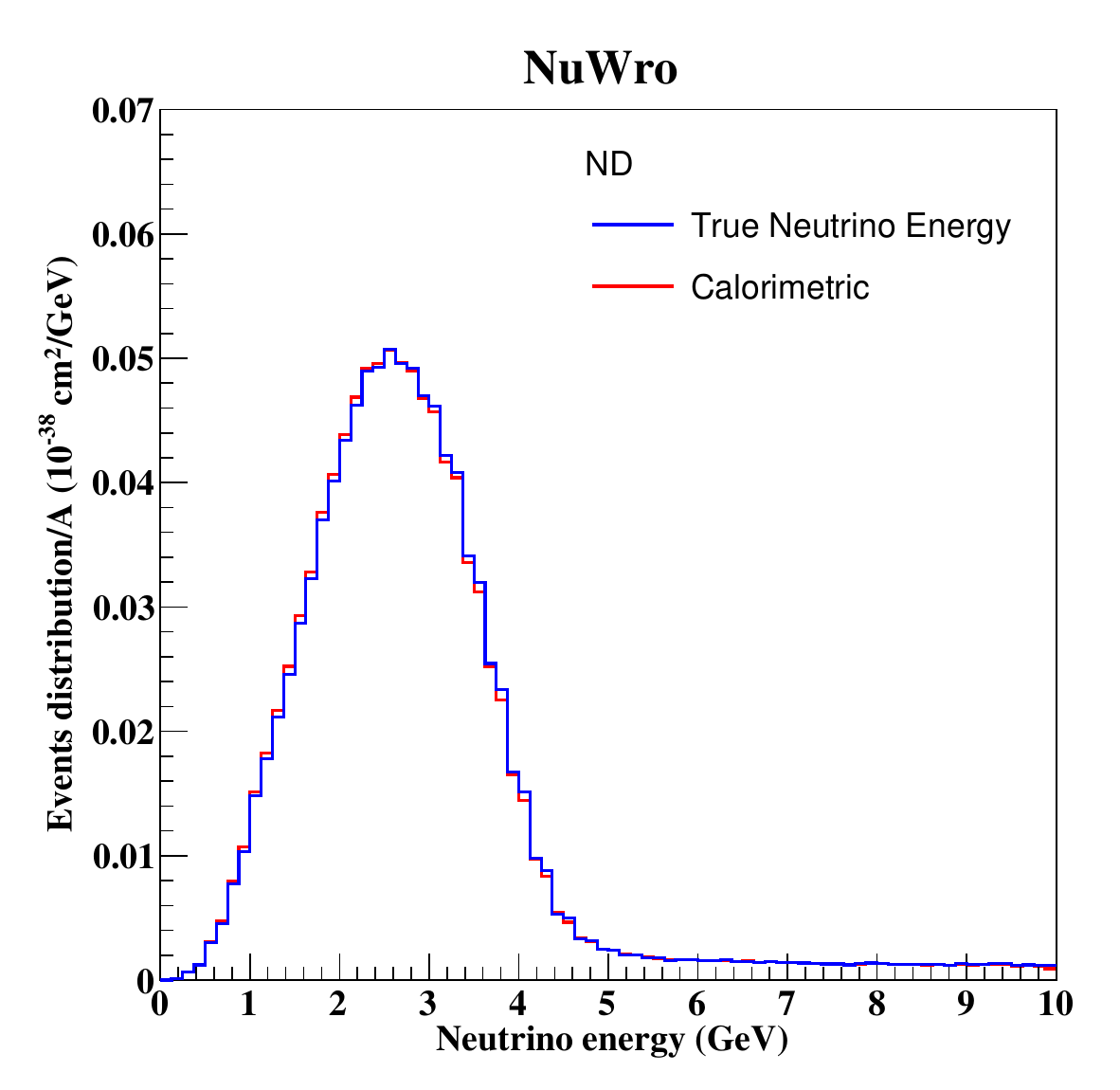}
		\vskip-3mm\caption{Event distribution at the ND as a function of reconstructed neutrino energy represented by red Calorimetric method, and as a function of true neutrino energy represented by blue lines for $\nu_{\mu}$-Ar (top) and $\bar{\nu_{\mu}}$-Ar (bottom) from GENIE and NuWro in the left and right panels, respectively.}
		\label{fig2}  
	\end{figure*}
	\begin{figure*}% figure* for wide figure, [h] [!] to change the placement
		\vskip1mm
		\includegraphics[scale=0.4]{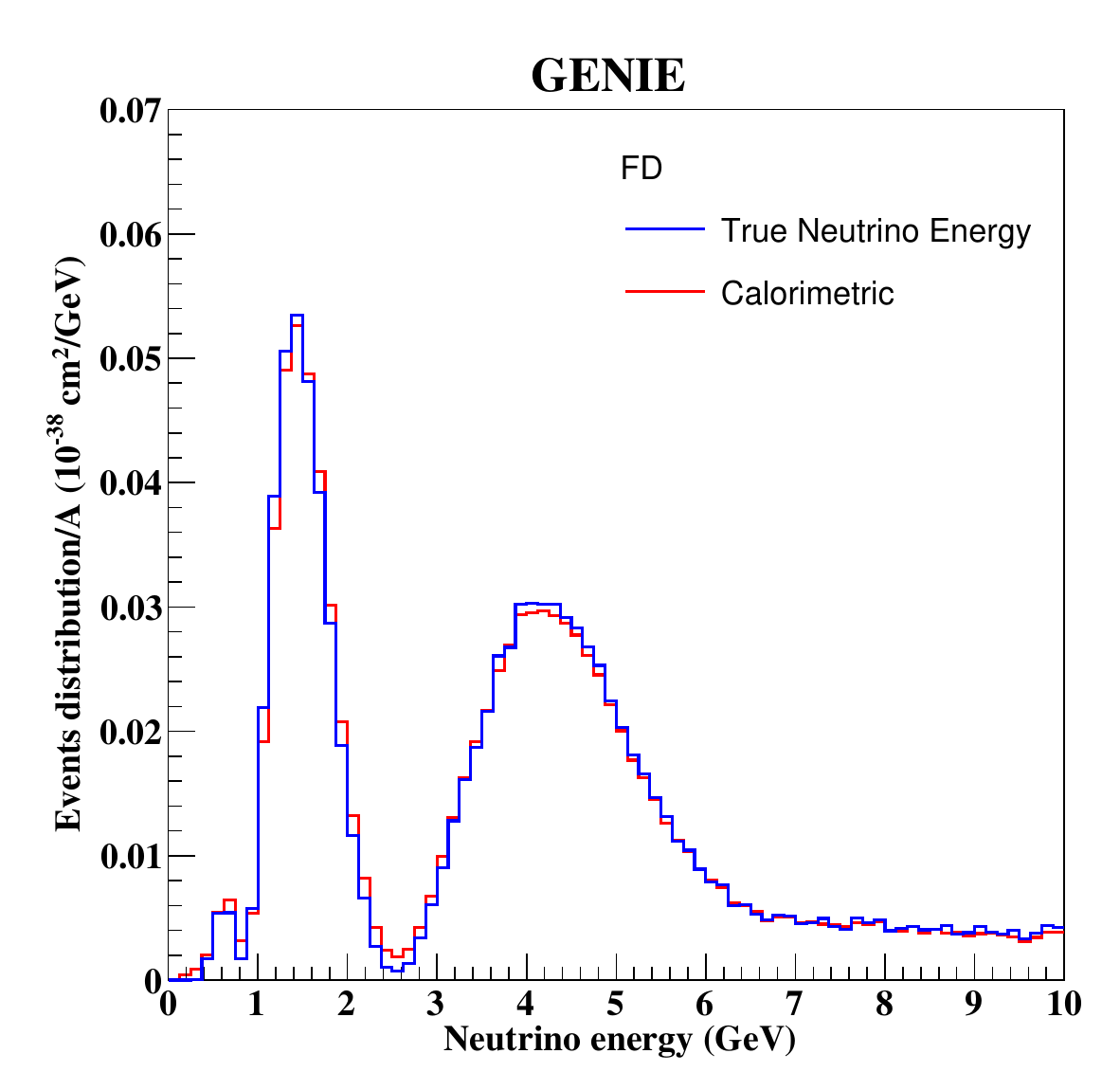} 
		\includegraphics[scale=0.4]{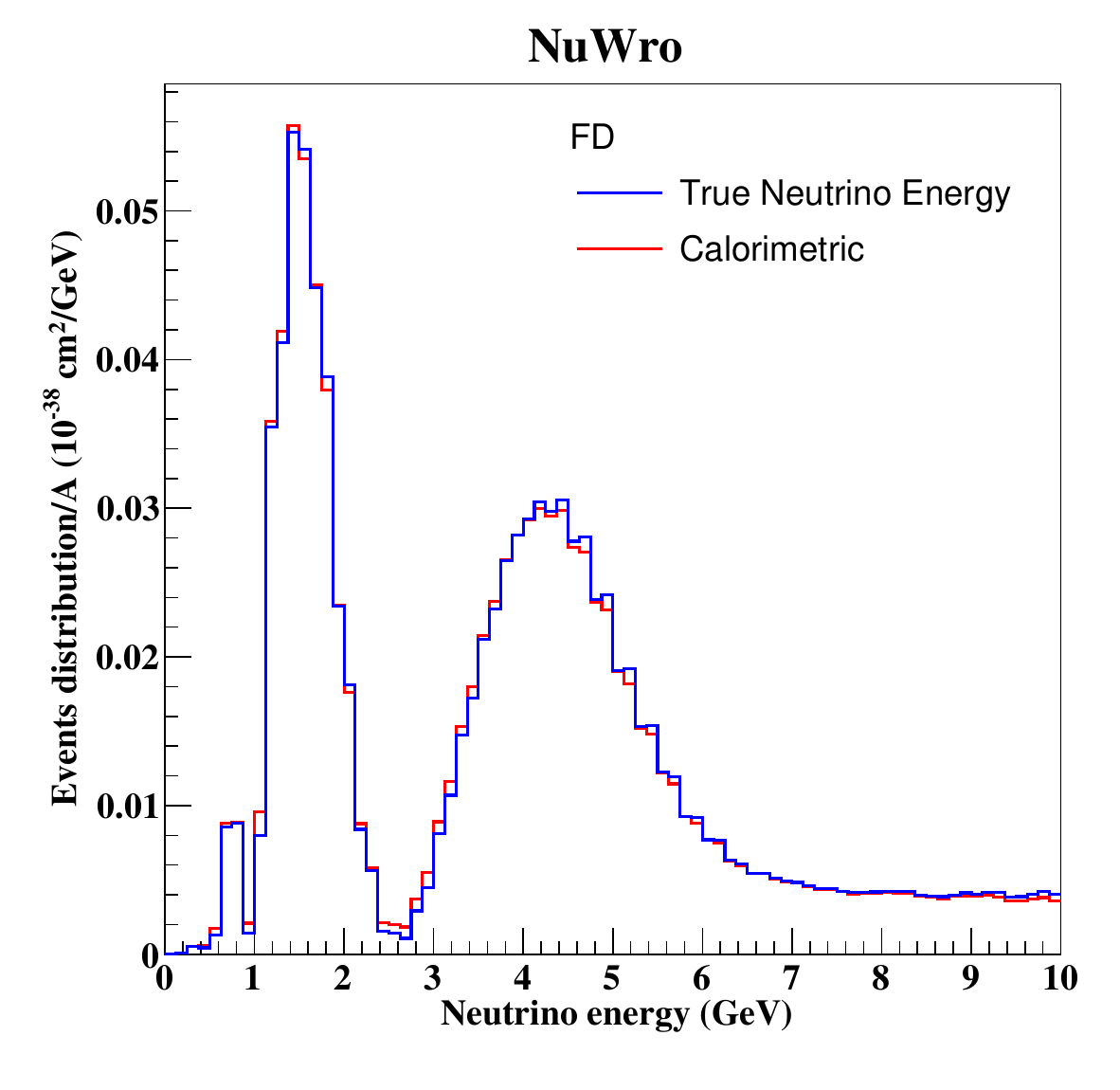}
		\includegraphics[scale=0.4]{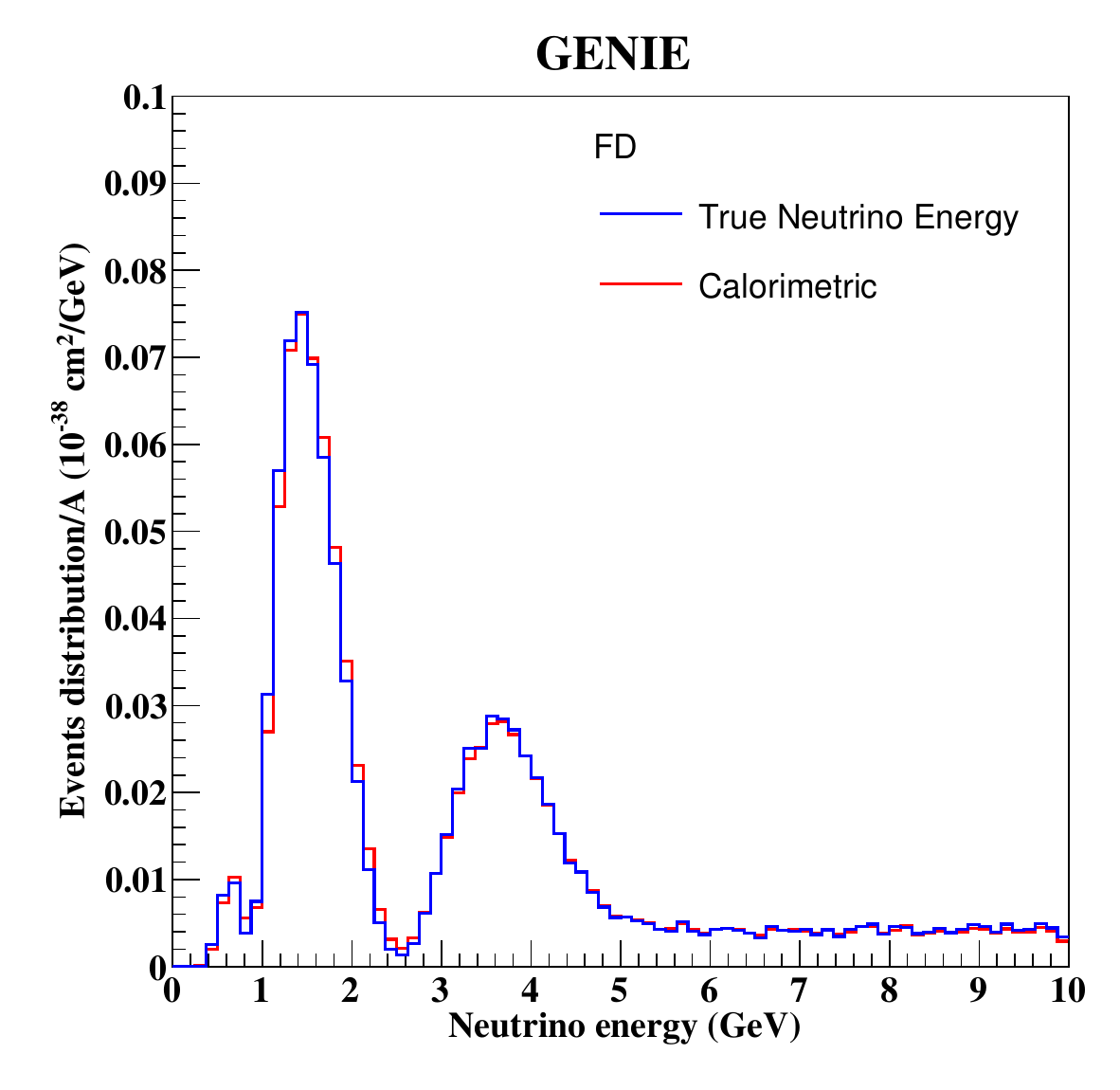} 	\includegraphics[scale=0.4]{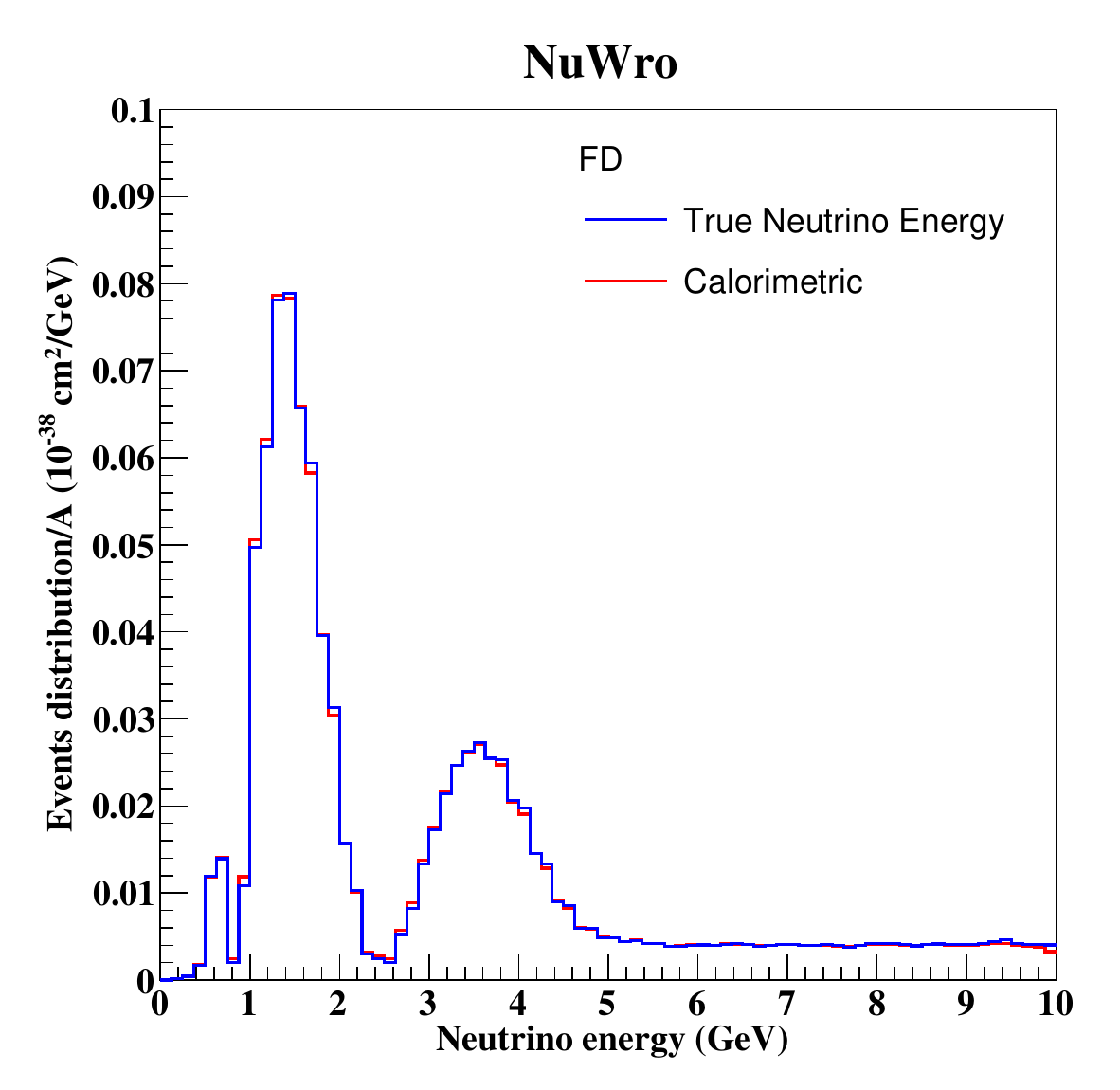}
		\vskip-3mm\caption{Event distribution at the FD as a function of reconstructed neutrino energy represented by red Calorimetric method, and as a function of true neutrino energy represented by blue lines for $\nu_{\mu}$-Ar (top) and $\bar{\nu_{\mu}}$-Ar (bottom) from GENIE and NuWro in the left and right panels, respectively.}
		\label{fig3}  
	\end{figure*}
	\begin{figure*}% figure* for wide figure, [h] [!] to change the placement
		\vskip1mm
		\includegraphics[scale=0.4]{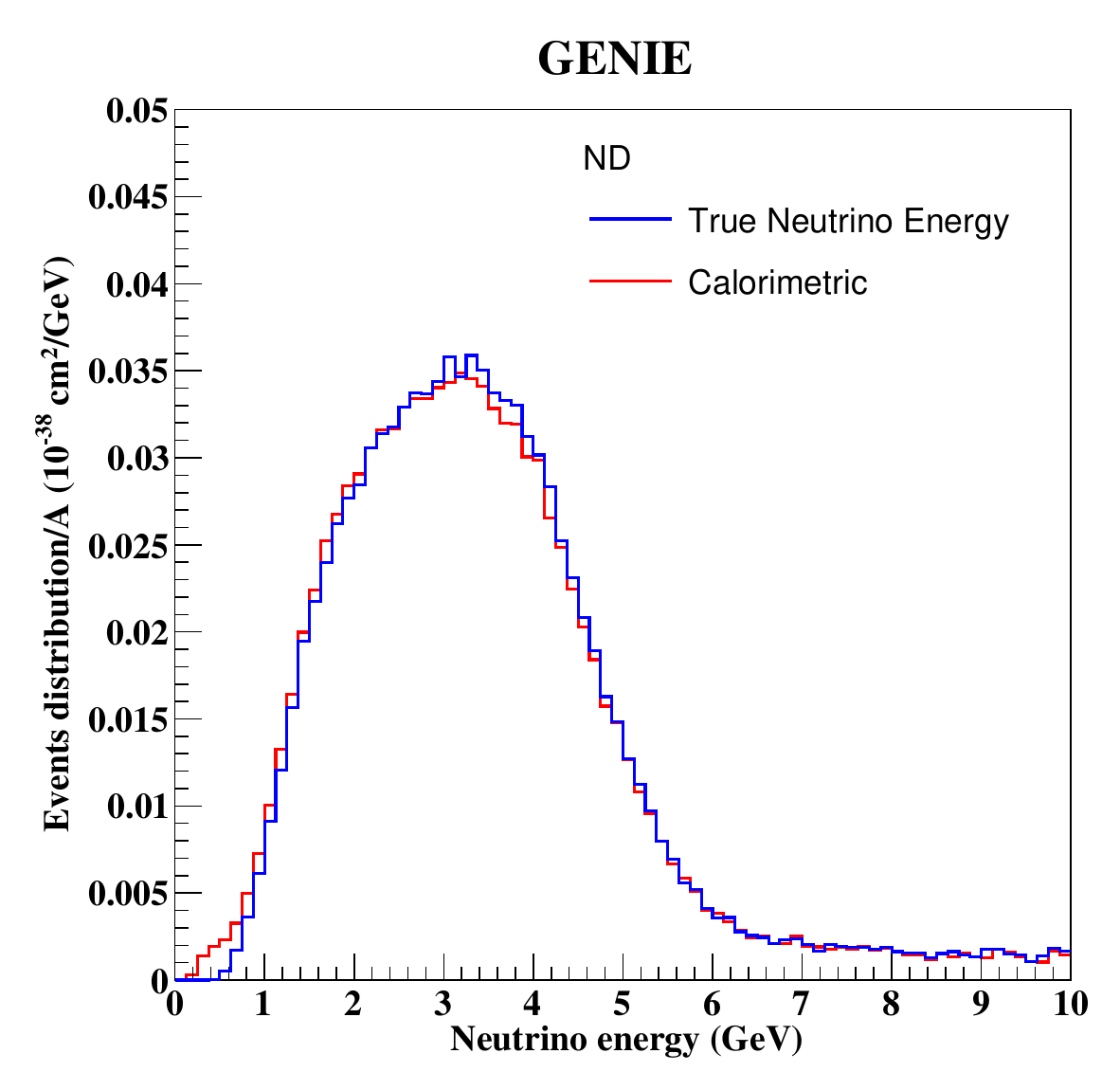} 	
		\includegraphics[scale=0.4]{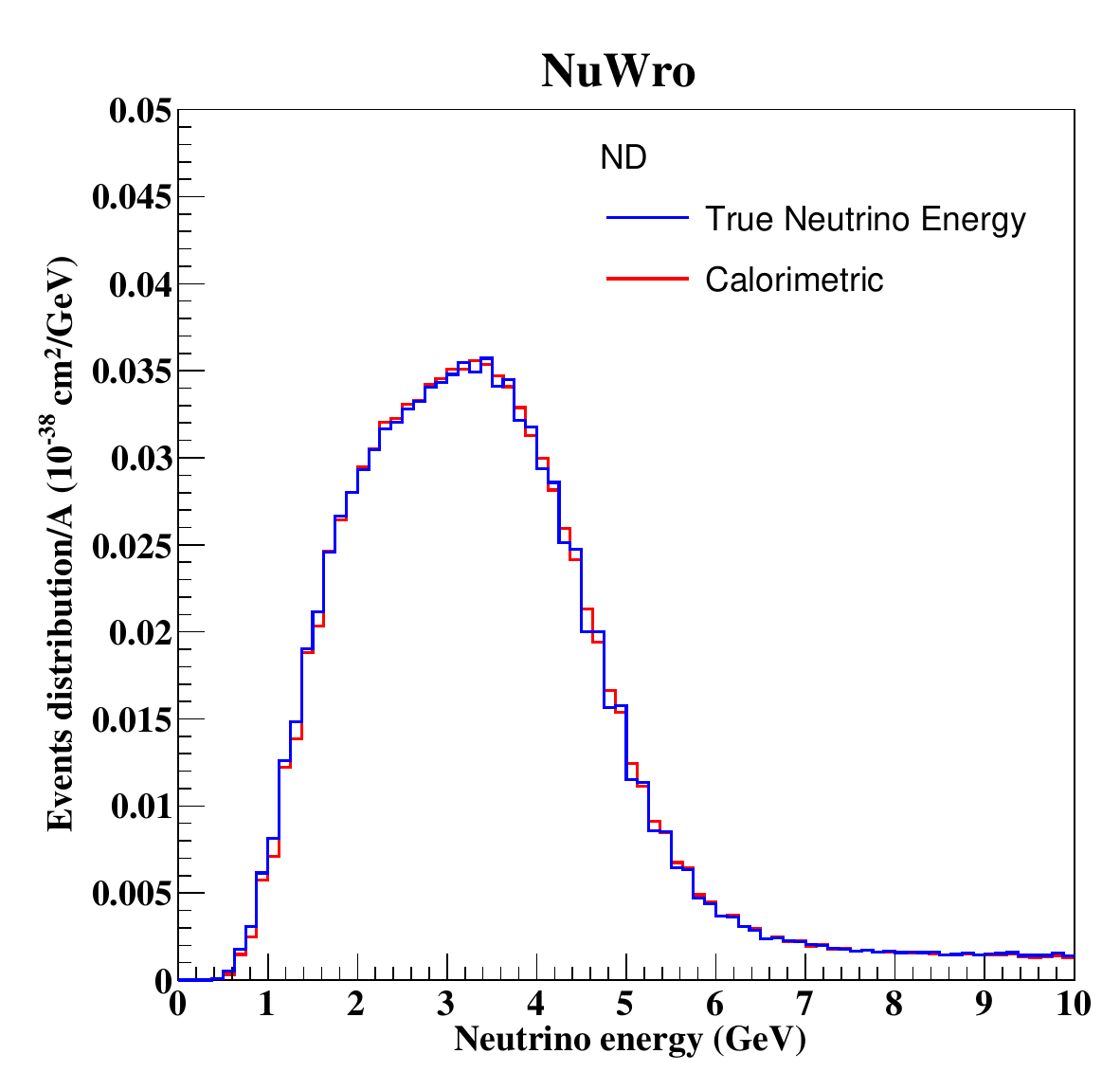}
		\includegraphics[scale=0.4]{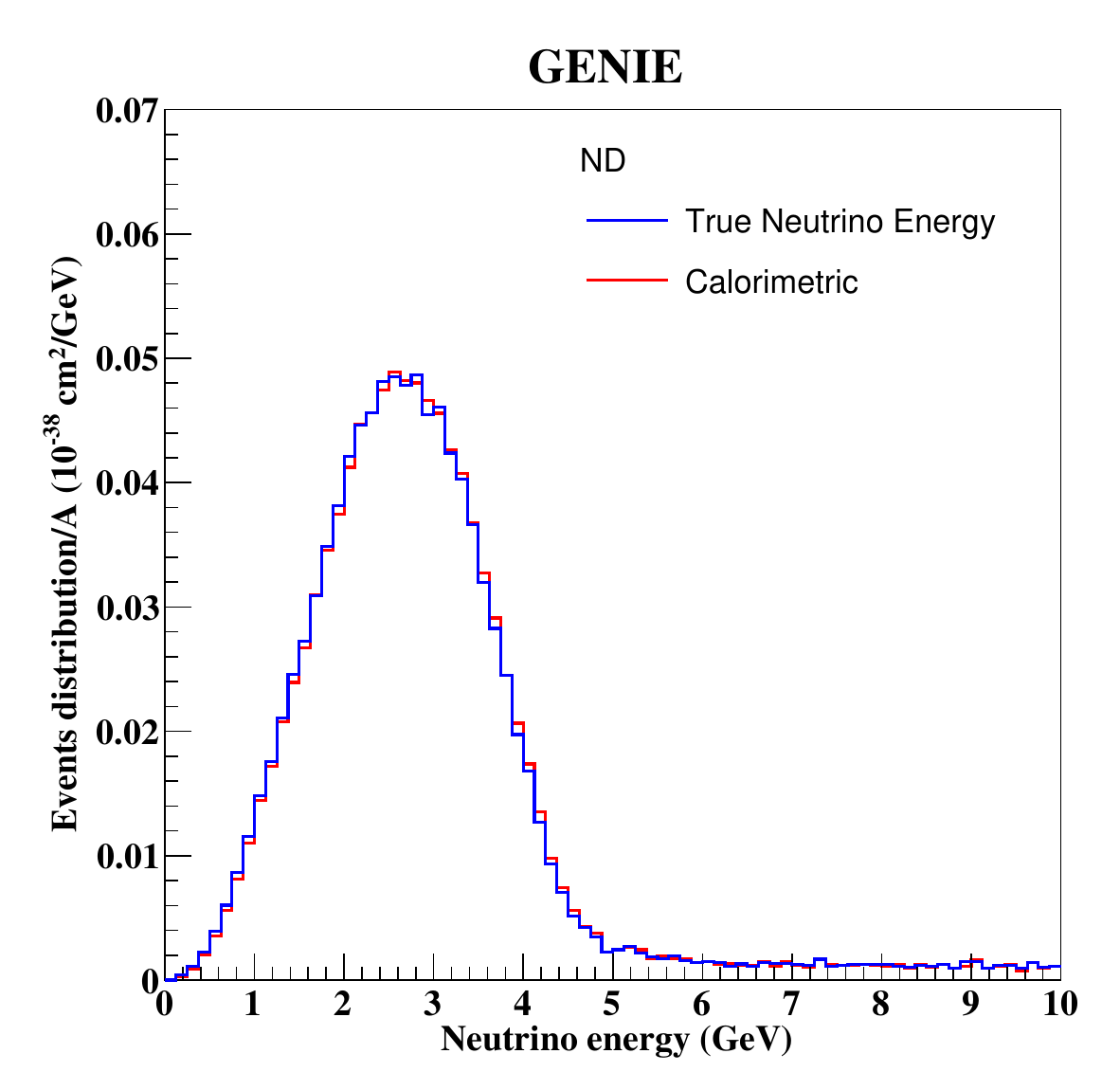} 	\includegraphics[scale=0.4]{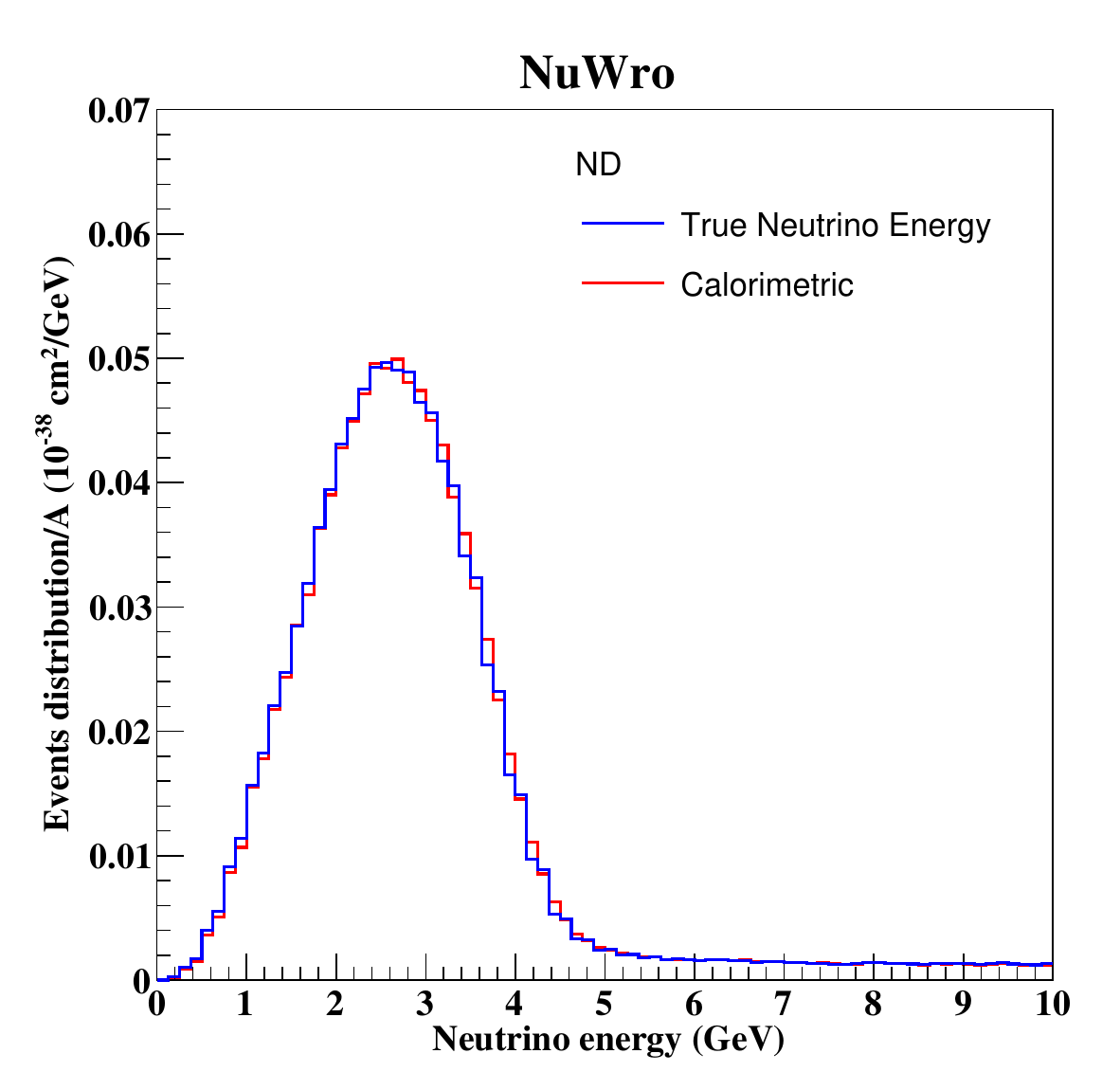}
		\vskip-3mm\caption{Event distribution at the ND as a function of reconstructed neutrino energy represented by red Calorimetric method, and as a function of true neutrino energy represented by blue lines for $\nu_{\mu}$-H (top) and $\bar{\nu_{\mu}}$-H (bottom) from GENIE and NuWro in the left and right panels, respectively.}
		\label{fig4}  
	\end{figure*}
	\begin{figure*}% figure* for wide figure, [h] [!] to change the placement
		\vskip1mm
			\includegraphics[scale=0.4]{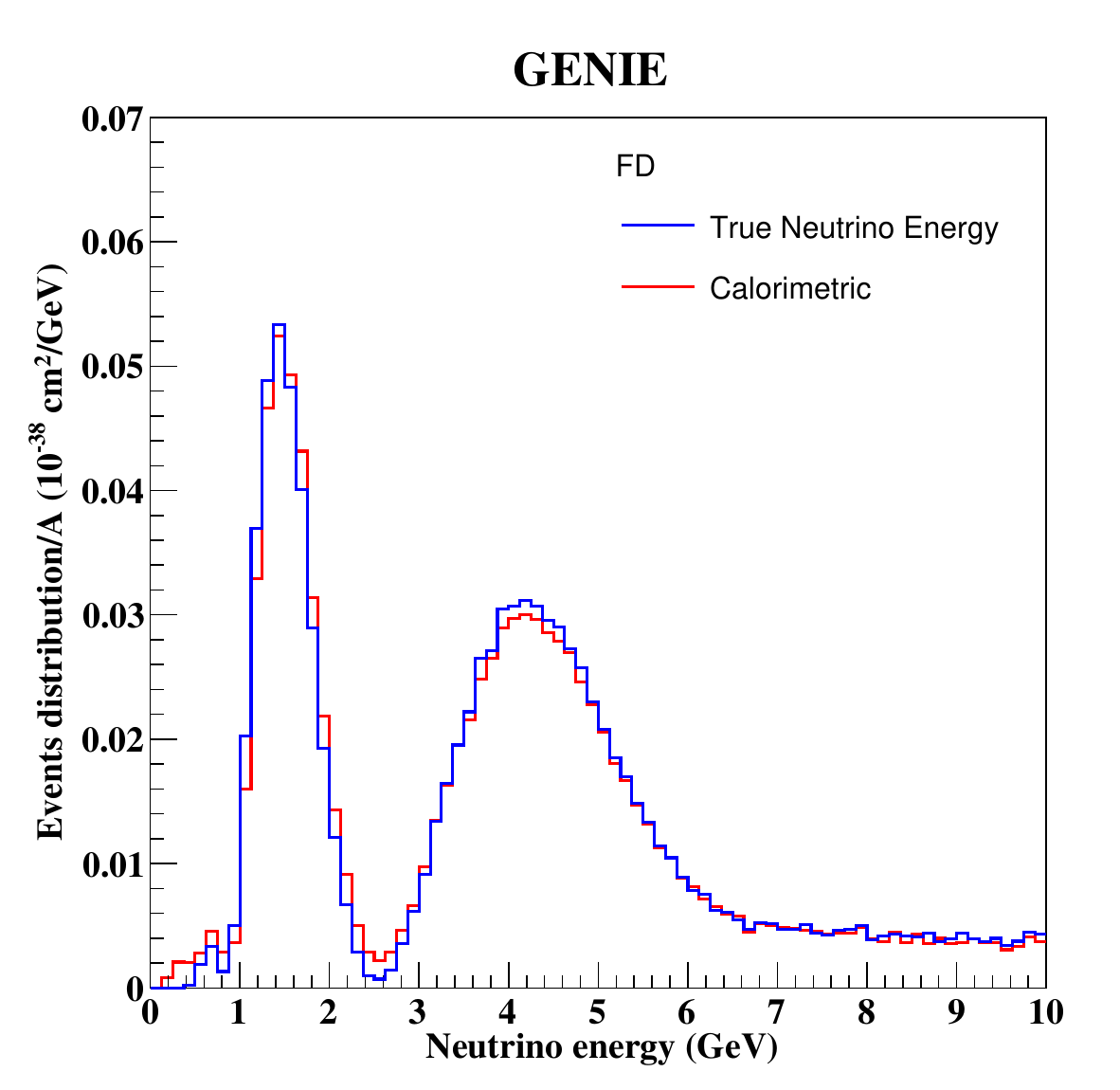} 	
		\includegraphics[scale=0.4]{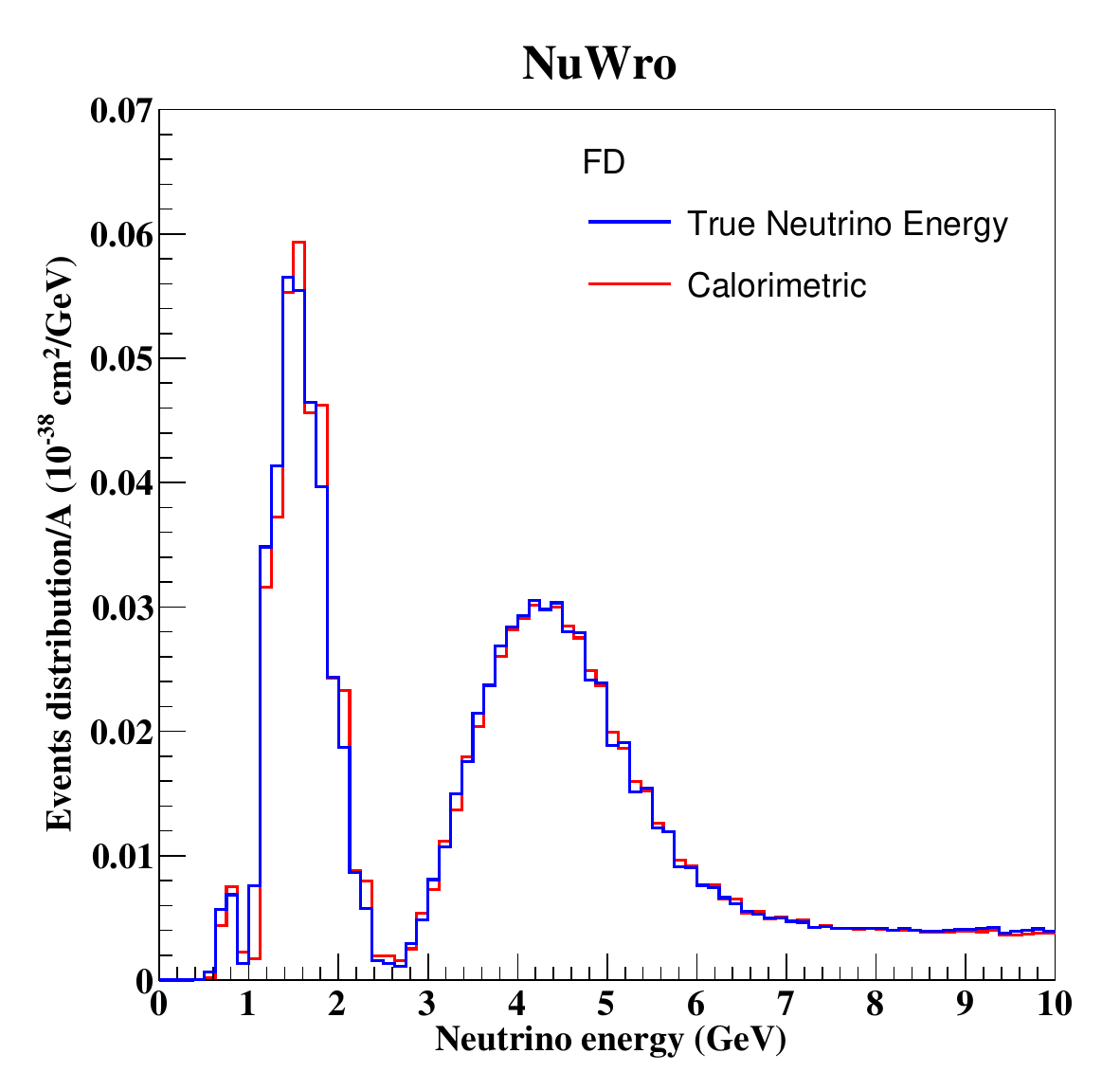}
		\includegraphics[scale=0.4]{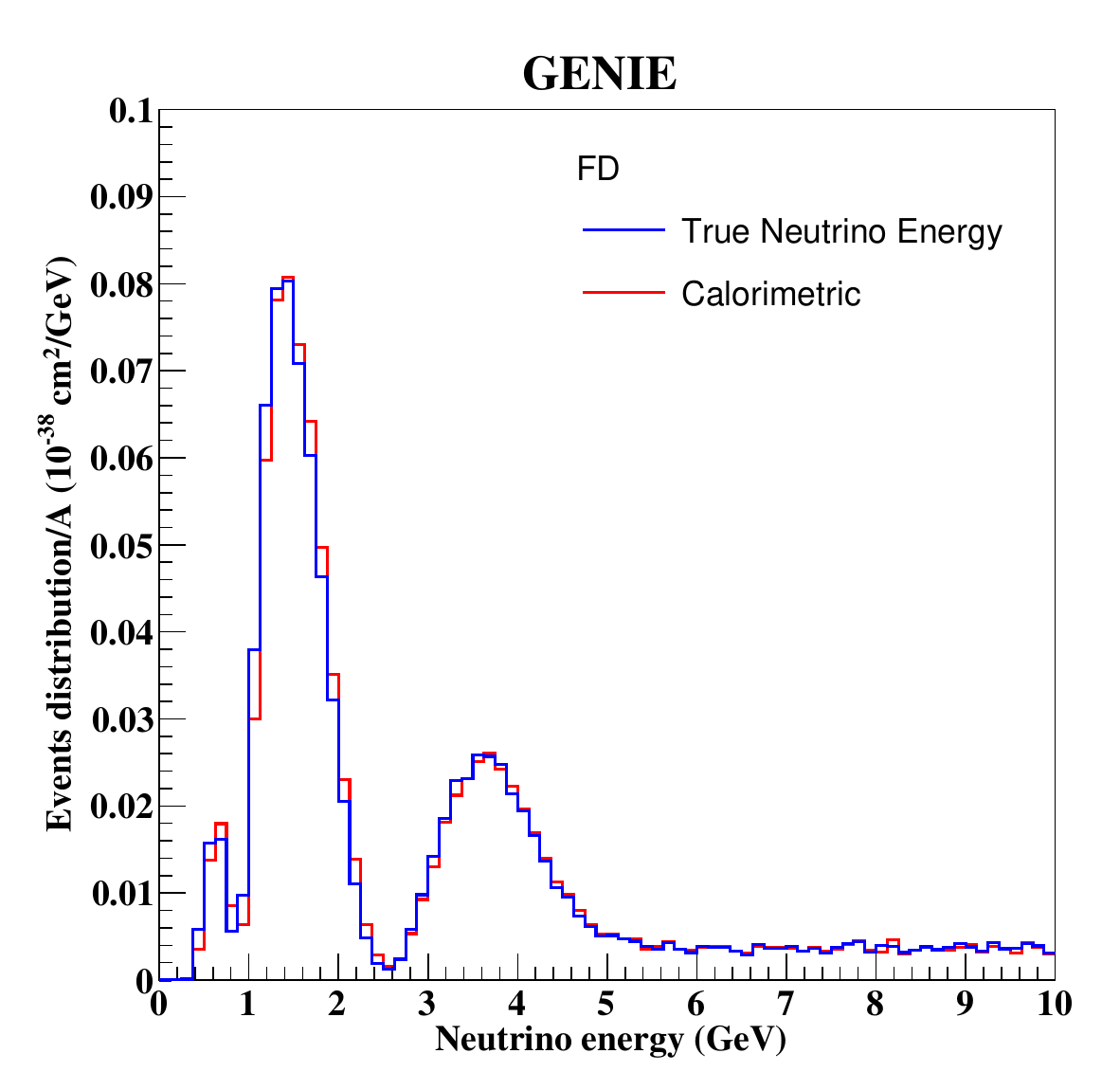} 	\includegraphics[scale=0.4]{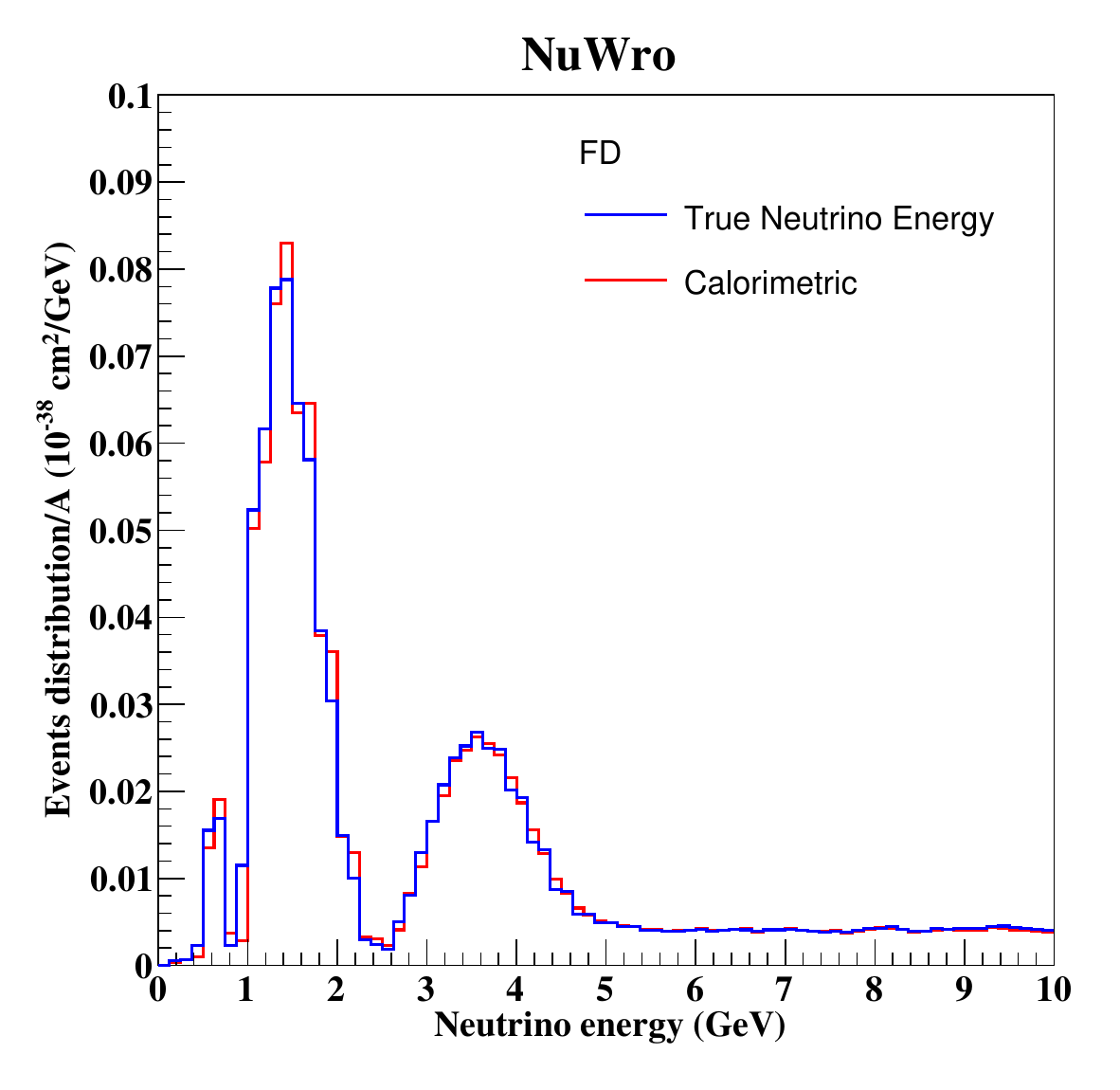}
		\vskip-3mm\caption{Event distribution at the FD as a function of reconstructed neutrino energy represented by red Calorimetric method, and as a function of true neutrino energy represented by blue lines for $\nu_{\mu}$-H (top) and $\bar{\nu_{\mu}}$-H (bottom) from GENIE and NuWro in the left and right panels, respectively.}
		\label{fig5}  
	\end{figure*}

	\begin{table}[!htp]
		\noindent\caption{Oscillation parameters considered in our work.}\vskip3mm\tabcolsep4.5pt
		
		\noindent{\footnotesize
			\begin{tabular}{|c | c| }
				\hline%
				\multicolumn{1}{|c}{\rule{0pt}{5mm}Parameter}%

				& \multicolumn{1}{|c|}{Central Value}\\[2mm]%
				\hline%
				\rule{0pt}{5mm} 	$\theta_{12}$(NO)           &   $0.5903$                                     \\
				$\theta_{13}$ (NO)              &   $0.150$                            \\
				$\theta_{23}$(NO)           &   $0.866$         \\
				$\delta_{CP}$               &   $90^{\circ}$          \\
				$\Delta m^{2}_{21}$ (NO)        &   7.39e-5$eV^{2}$                         \\
				$\Delta m^{2}_{32}$ (NO)    &  2.451e-3 $eV^{2}$     \\
				$\Delta m^{2}_{31}$(NO)     &  2.525e-3  $eV^{2}$     \\
				$\rho$                      &  2.848 $g$ $cm^{-3}$    \\
				[2mm]%
				\hline
			\end{tabular}
		}
		\label{table1}
	\end{table}
	
	The oscillation probability used here is represented by equation \ref{eq1} and the true set of values of the oscillation parameters \cite{4DUNE} considered in this analysis are presented in Table \ref{table1}.
	One can see that the oscillation parameters and neutrino energy are involved in the oscillation probabilities equation. The uncertainties in the measured neutrino energy directly translate into uncertainties in oscillation parameters. The oscillation probability of $\nu_{\mu}\rightarrow \nu_{e}$ channel through matter for the three-flavor scenarios and constant density approximation is written as, \cite{4DUNE}

	\begin{equation} \label{eq1}
	\begin{split}
	P\left( \overset{(-)}{\nu_{\mu}}\rightarrow\overset{(-)}{\nu_{e}}\right)= & \sin^2\theta_{23}\sin^22\theta_{13}\dfrac{\sin^2 (\Delta_{31} -aL )}{(\Delta_{31} -aL )^2}\Delta_{31}^2 \\
	&+  \sin 2\theta_{23} \sin 2\theta_{13} \sin 2\theta_{12} \dfrac{\sin( \Delta_{31} -aL )}{(\Delta_{31} -aL )} \Delta_{31} \\ & 
	\times \dfrac{\sin(aL)}{aL} \Delta_{21} \cos(\Delta_{31}  \pm \delta_{CP}) \\ &+ \cos ^2 \theta_{23} \sin^2 2\theta_{12} \dfrac{\sin^2aL}{(aL)^2} \Delta_{21},
	\end{split}
	\end{equation}
	
	where
	\begin{equation}
	a= \dfrac{G_F N_e}{\sqrt{2}} \approx  \pm \dfrac{1}{3500 \space km}\left(\dfrac{\rho}{ {3.0} {g/cm^3}}\right) ,
	\end{equation}
	
	$G_F$ is the Fermi constant, $N_e$ denotes the number density of electrons in the Earth\textsc{\char13}s crust, $\Delta_{ij}= $ 1.267 $\Delta m^2_{ij}L/E_\nu$, $L =1285$ is the distance from the neutrino source to the detector in km, and $E_{\nu}$ stands for neutrino energy in GeV. Both $\delta_{CP}$ and $a$ terms are positive for $\nu_{\mu} \rightarrow \nu_{e}$ and negative for $ \overset{-}{\nu_{\mu}} \rightarrow \overset{-}{\nu_{e}} $ oscillations.
	
	In our analysis, we have used the calorimetric method for energy reconstruction. The calorimetric method of neutrino energy reconstruction demands information of all the detectable final state particles on an event by event basis i.e. total of the energy deposited by all reaction products in the detector and thus permitting for precise reconstruction of neutrino energy. It is appropriate to all types of events, unlike the kinematic method which is only applicable for QE event reconstruction. Still, this method gives rise to challenges in the way of true neutrino energy reconstruction. The main challenges are the precise reconstruction of all produced hadrons, where neutrons are assumed to escape detection completely. Thus, the calorimetric method relies on the capability of detector design and performance in reconstructing final states.  The neutrino reconstructed energy $ E^{Calori}_{\nu} $  \cite{2i} in CC processes resulting in the knockout of $n$ nucleons and production of $m$ mesons, can be simply determined using the calorimetric method, 
	
	\begin{equation}
	E^{Calori}_{\nu} =   E_{lep}+ \epsilon_{nuc}+\sum_{a}(E_{n_{a}}-M)+ \sum_{b}E_{m_{b}}
	\end{equation}
	
	Where $E_{lep}$ is the  energy of outgoing lepton, $ E_{n_{a}}$ denotes the energy of the $a^{th}$ knocked out nucleon, $ \epsilon_{nuc} $ denotes the corresponding separation energy  and the single-nucleon separation energy is fixed to 34 MeV. $E_{m_{b}}$ stands for the energy of $b^{th}$ produced meson, and $M$ is the mass of nucleon. 
	\section{Results and discussion}
	\label{sec4}

	\begin{figure*}% figure* for wide figure, [h] [!] to change the placement
		\vskip1mm
		\includegraphics[scale=0.4]{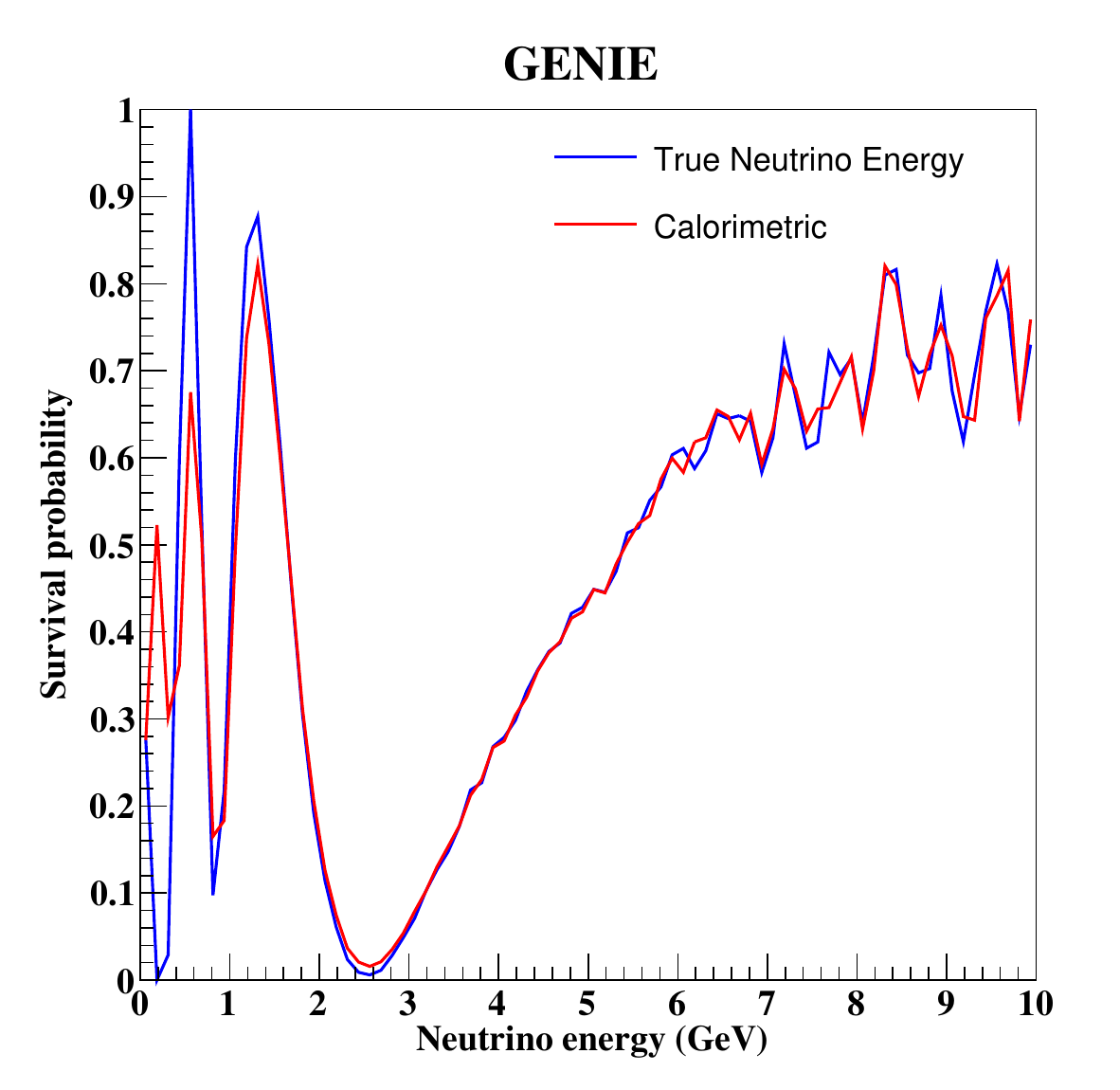} 	
		\includegraphics[scale=0.4]{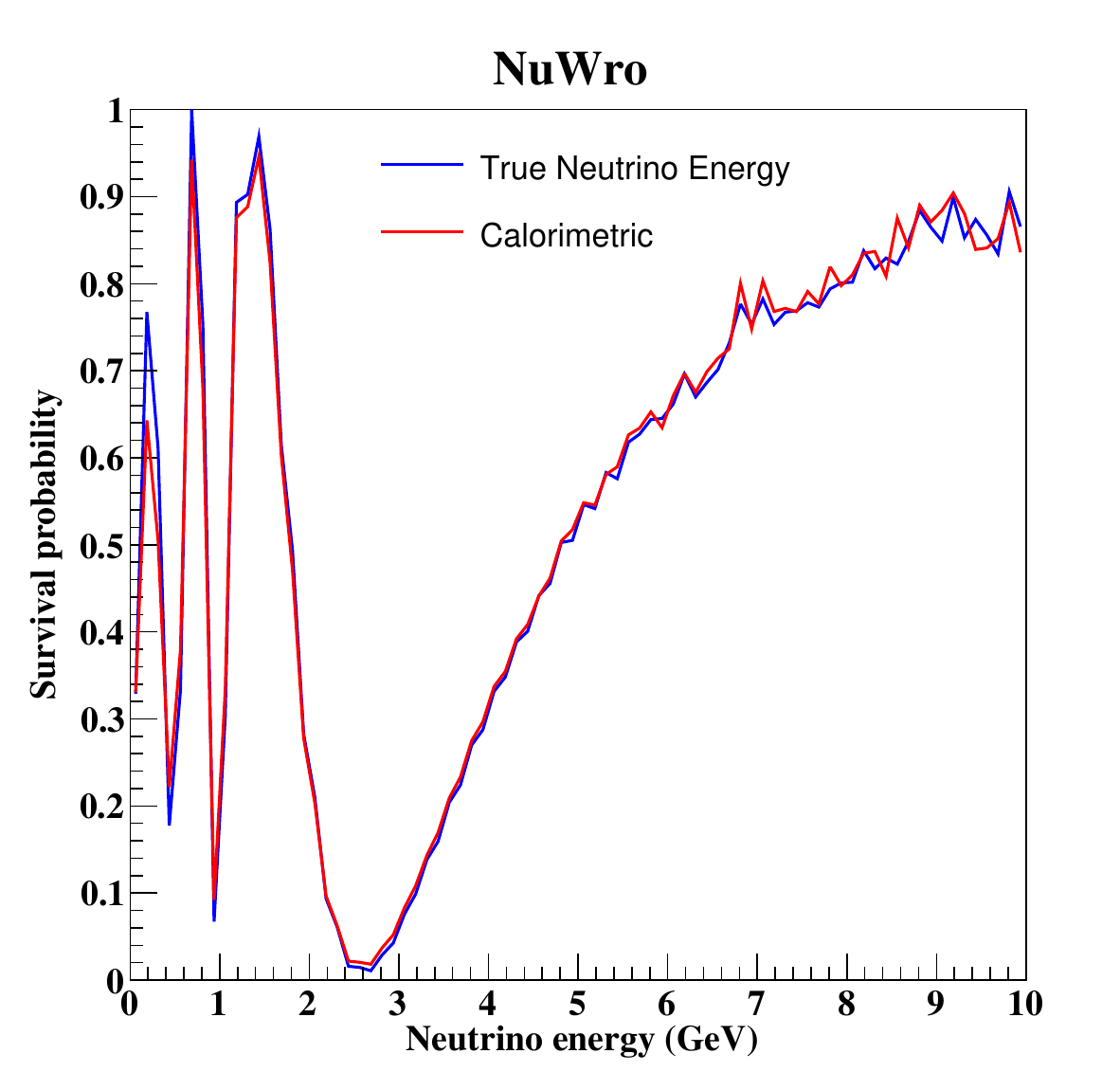}
		\includegraphics[scale=0.4]{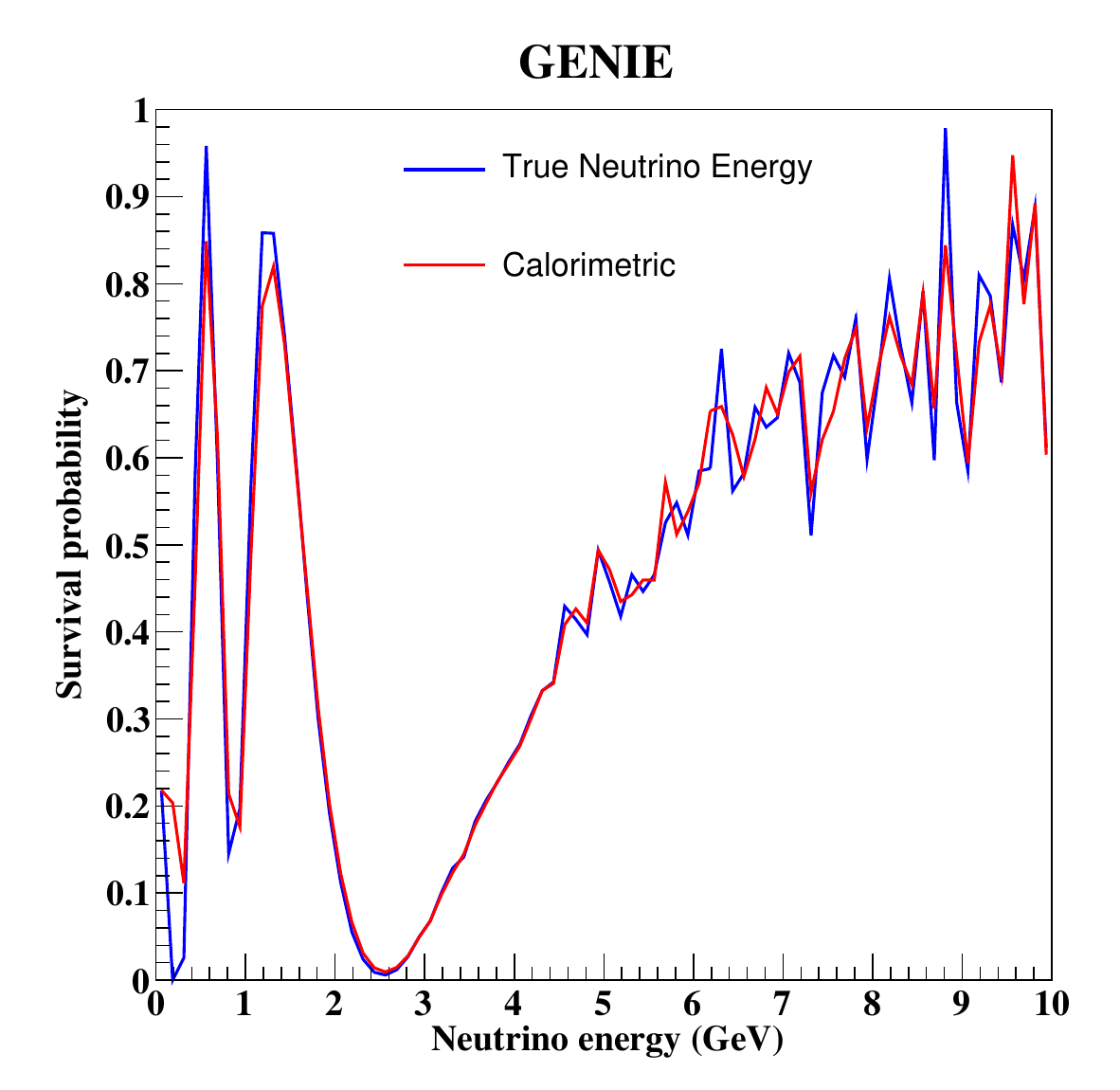}
				\includegraphics[scale=0.4]{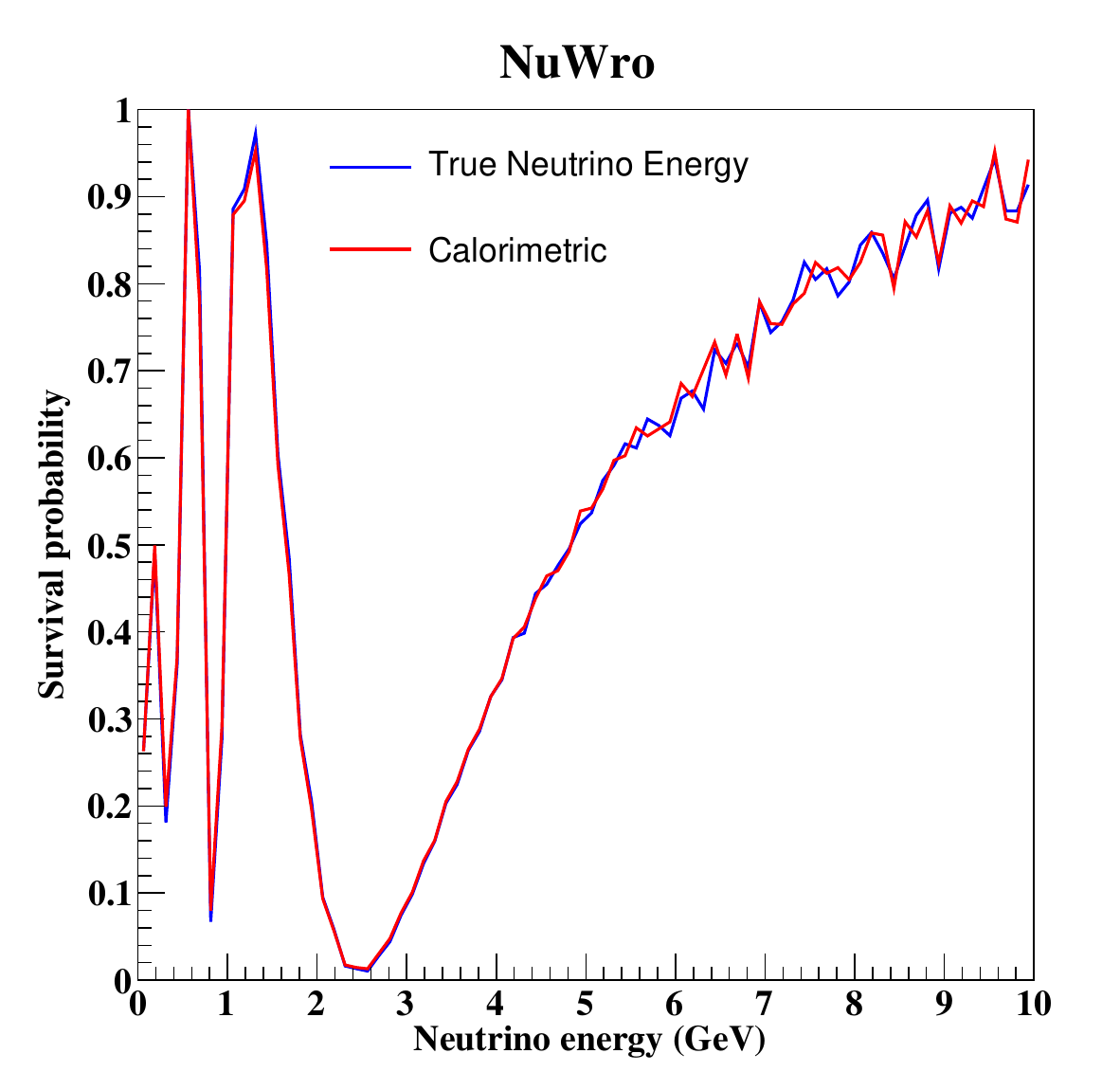}  
		\vskip-3mm\caption{Muon (anti)neutrino survival probability as a function of reconstructed neutrino energy represented by red Calorimetric method, and as a function of true neutrino energy represented by blue lines for $\nu_{\mu}$-Ar (top) and $\bar{\nu_{\mu}}$-Ar (bottom) from GENIE and NuWro in the left and right panels, respectively.}
		\label{fig6}  
	\end{figure*}

	\begin{figure*}% figure* for wide figure, [h] [!] to change the placement
		\vskip1mm
		\includegraphics[scale=0.4]{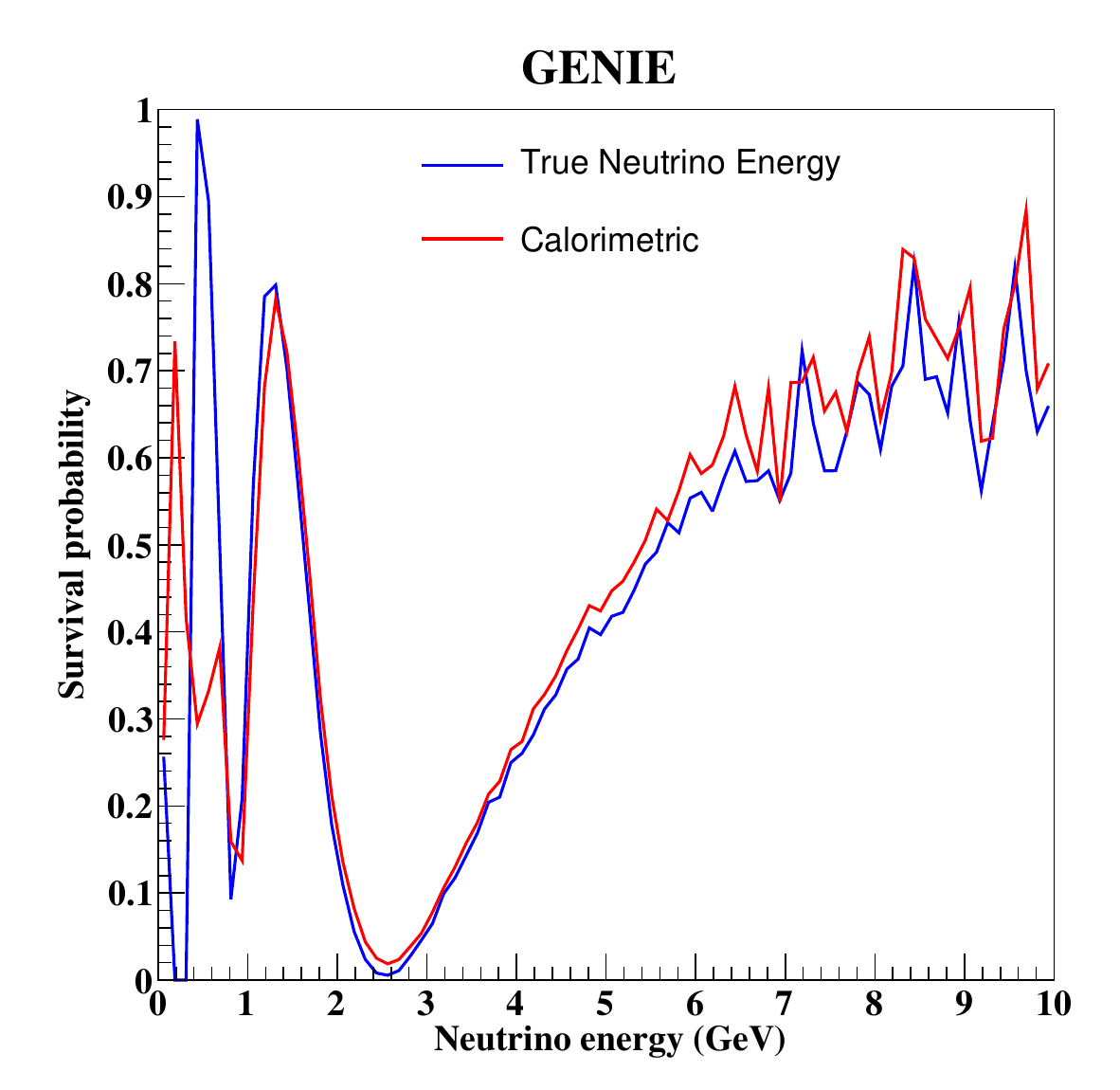} 	
		\includegraphics[scale=0.4]{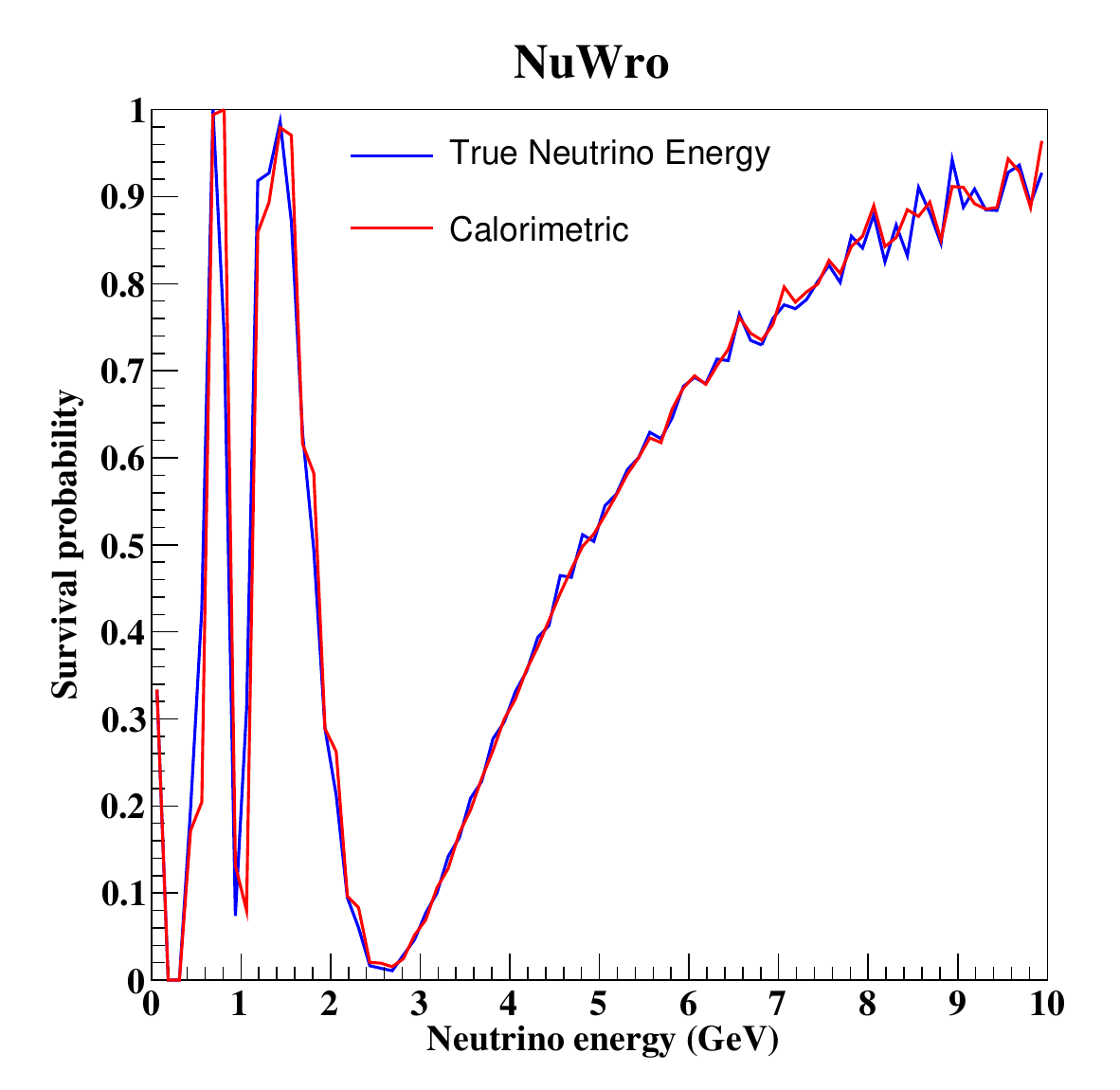}
		\includegraphics[scale=0.4]{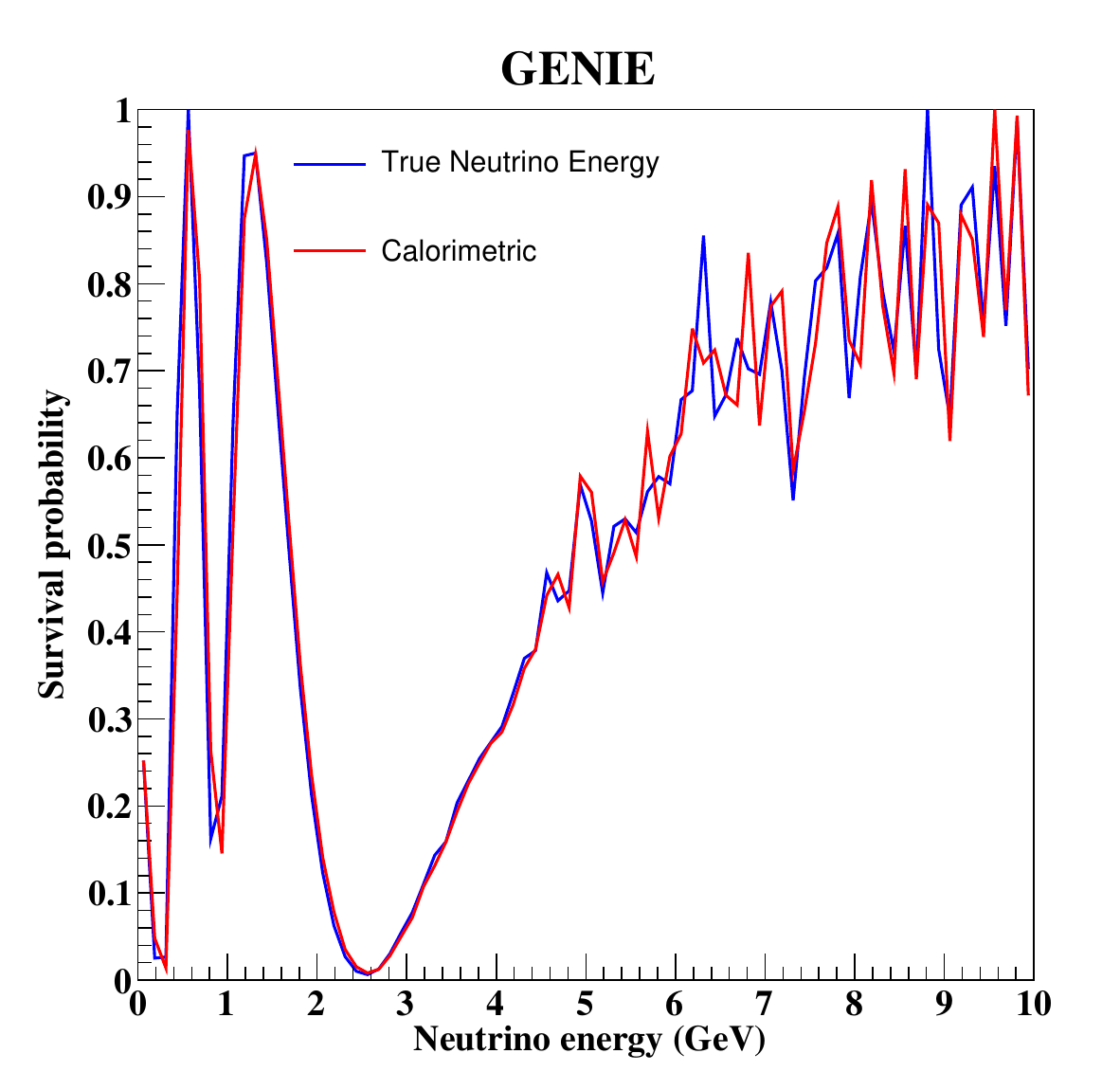} 	
		\includegraphics[scale=0.4]{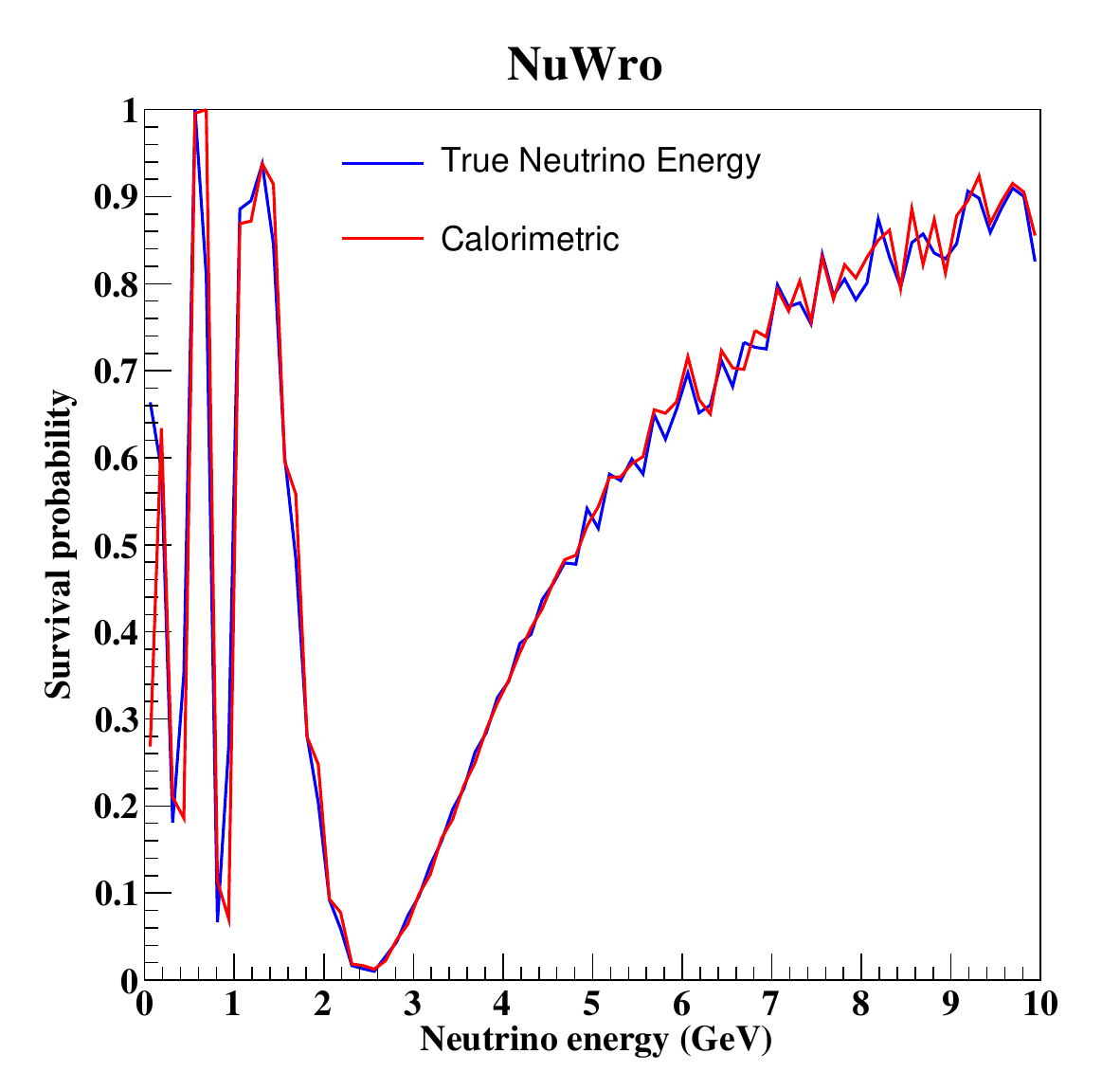}
		\vskip-3mm\caption{Muon (anti)neutrino survival probability as a function of reconstructed neutrino energy represented by red Calorimetric method, and as a function of true neutrino energy represented by blue lines for $\nu_{\mu}$-H (top) and $\bar{\nu_{\mu}}$-H (bottom) from GENIE and NuWro in the left and right panels, respectively.}
		\label{fig7}  
	\end{figure*}

	\begin{figure*}% figure* for wide figure, [h] [!] to change the placement
		\vskip1mm
		\includegraphics[scale=0.4]{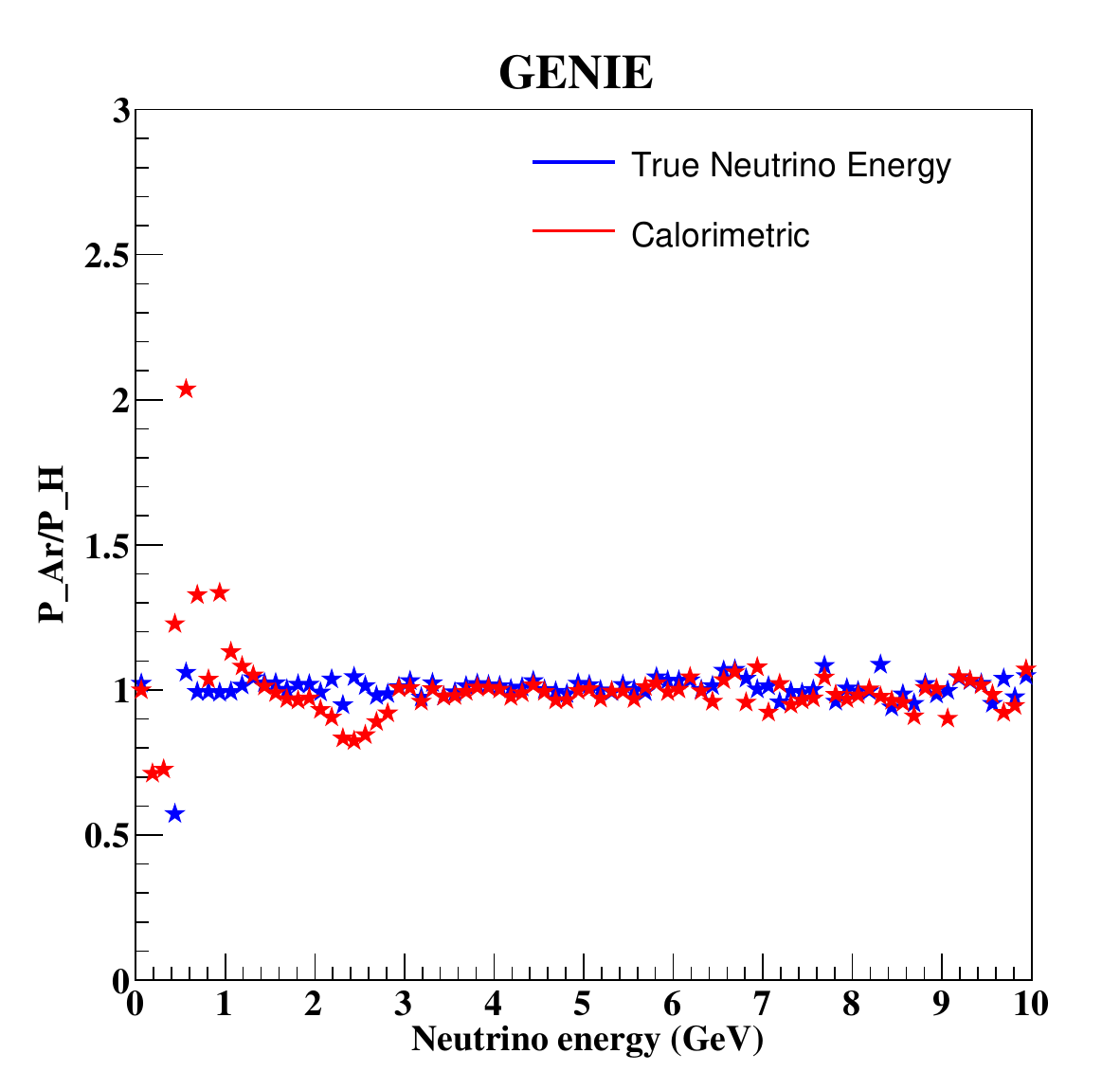} 	
		\includegraphics[scale=0.4]{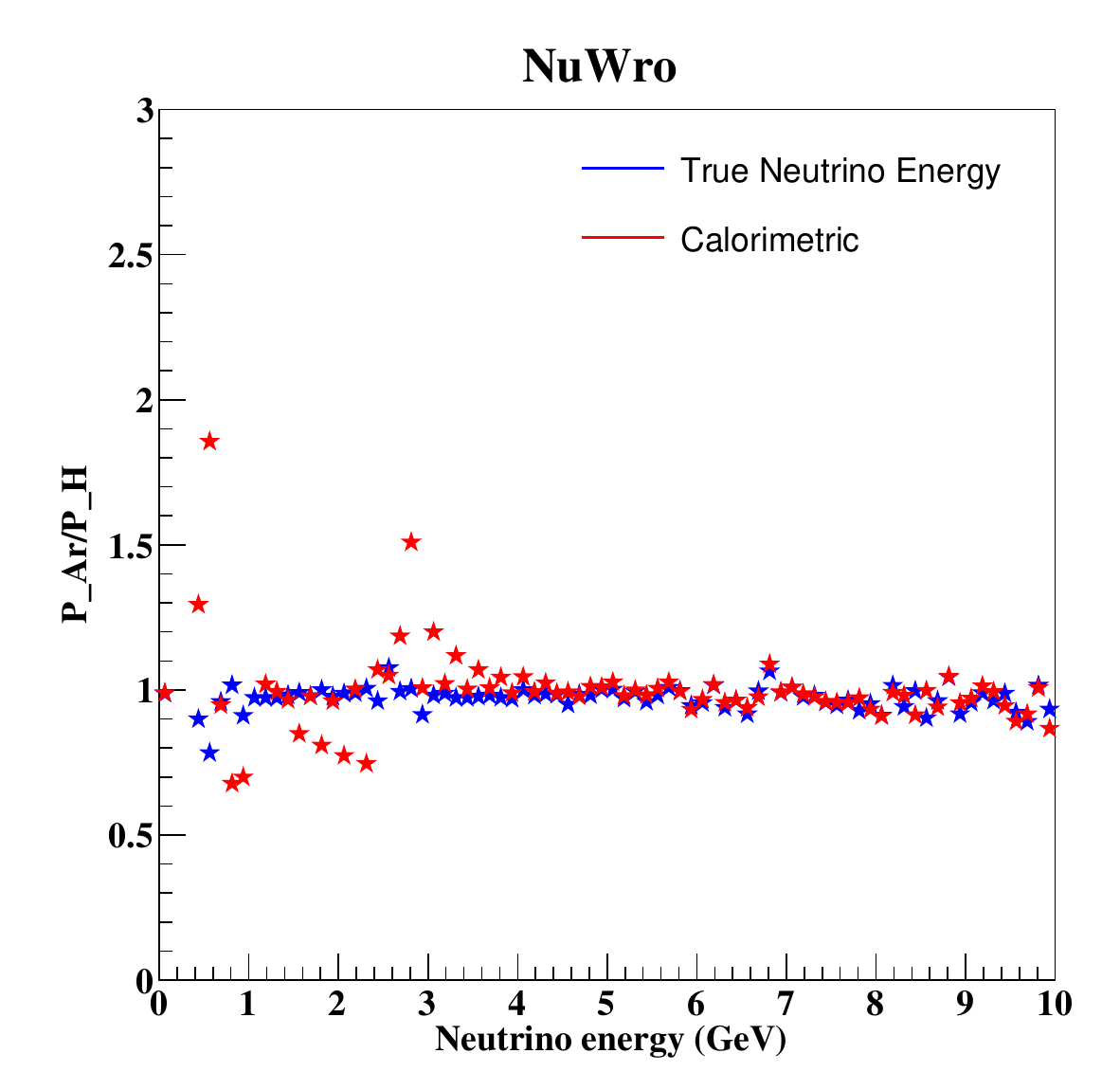}
		\includegraphics[scale=0.4]{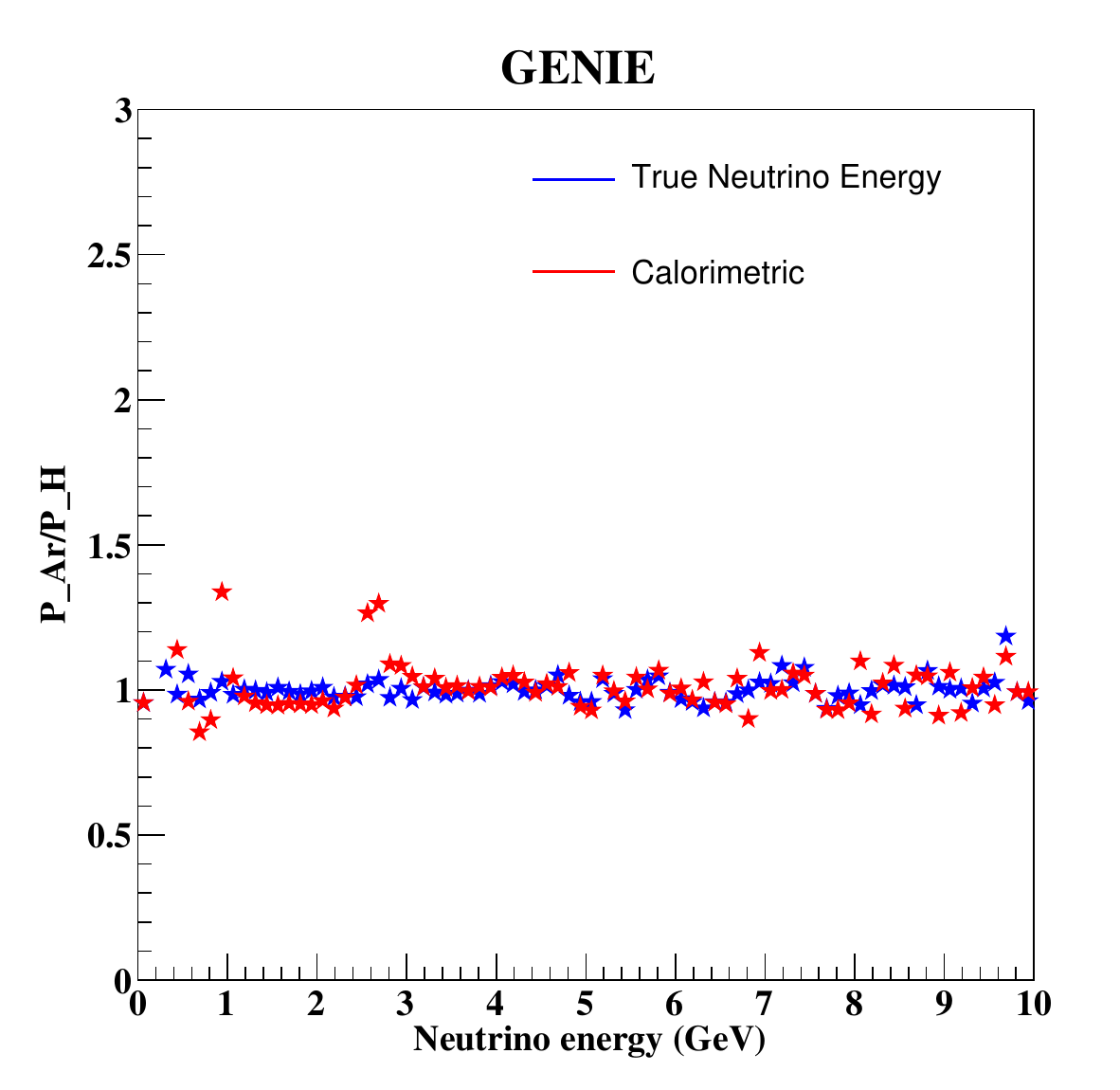} 	
		\includegraphics[scale=0.4]{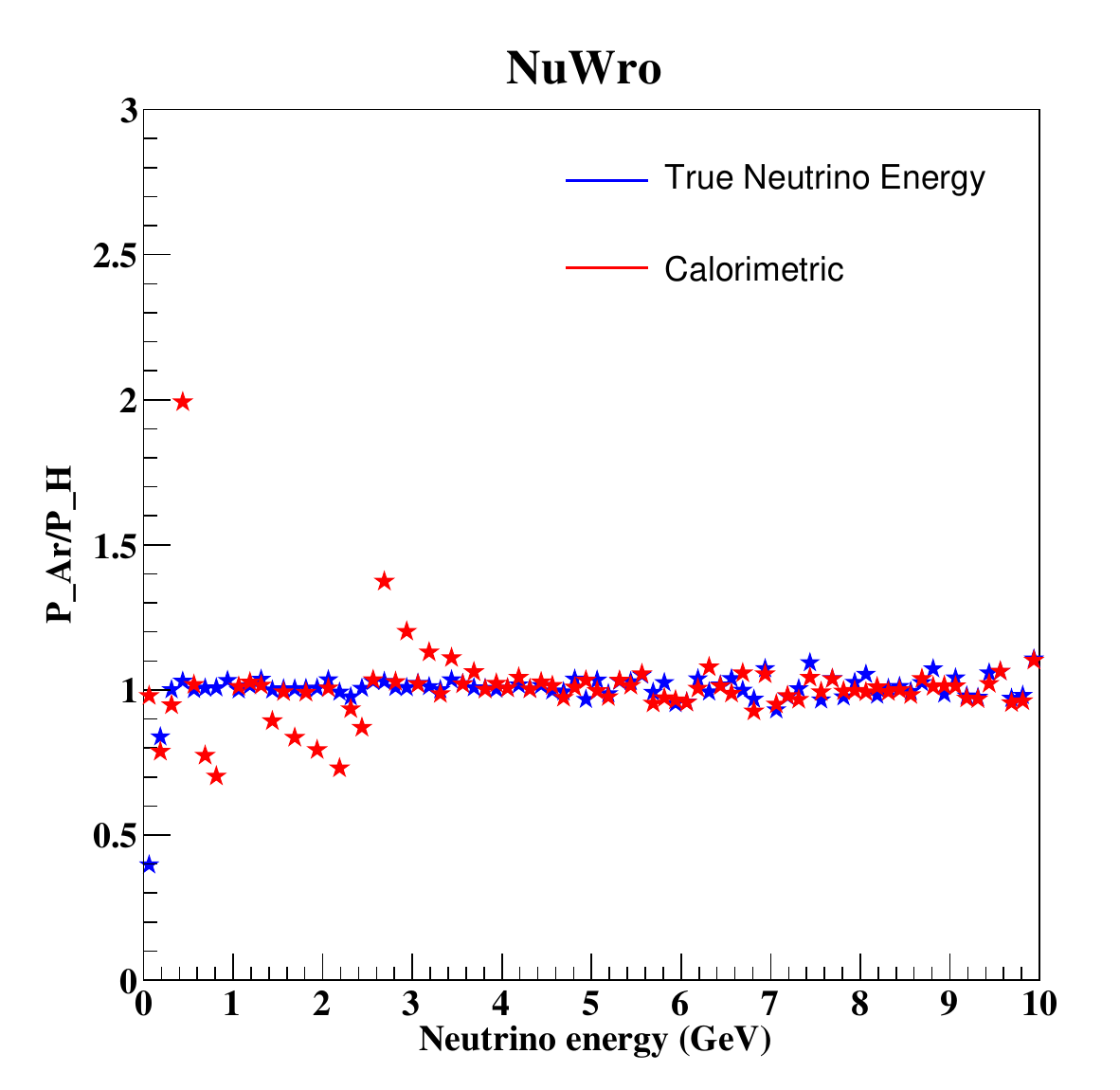}
		\vskip-3mm\caption{Ar/H Ratio of Muon (anti)neutrino survival probability as a function of reconstructed neutrino energy represented by red Calorimetric method, and as a function of true neutrino energy represented by blue lines for $\nu_{\mu}$ (top) and $\bar{\nu_{\mu}}$ (bottom) from GENIE and NuWro in the left and right panels, respectively.}
		\label{fig8}  
	\end{figure*}

	Figures \ref{fig2} and \ref{fig3}, represent the   event distribution  as a function of neutrino energy for $\nu$-Ar (upper panel) and  $ \bar{\nu}$-Ar (lower panel) at  ND, without oscillations and at the FD, with oscillations, respectively  using GENIE (left panel) and NuWro (right panel). We notice that the reconstructed neutrino energy is very well reconstructed using the calorimetric method for both the generators. 
	
	Figures \ref{fig4} and \ref{fig5}, represent the   event distribution  as a function of neutrino energy for $\nu$-H (upper panel) and  $ \bar{\nu}$-H (lower panel) at  ND, without oscillations and at the FD, with oscillations, respectively  using GENIE (left panel) and NuWro (right panel). We again notice that the reconstructed neutrino energy is very well reconstructed using the calorimetric method for both the generators.

	The oscillation probability can be calculated by dividing the number of events at FD by the number of events at ND \cite{mosel}. Figure \ref{fig6} represents the Muon (anti)neutrino  survival probability  for $\nu_{\mu}$-Ar (top) and $ \bar{\nu}$-Ar (bottom) as obtained from GENIE (left) and NuWro (right) and Figure \ref{fig7} shows same for H.  The Muon (anti)neutrino survival probability as calculated  from Calorimetric method  is represented by red line and the blue line represents the same for True neutrino energy.
	
	In order to measure the impact of systematic uncertainties arising due to nuclear effects in neutrino energy reconstruction, we have taken the ratio of Muon (anti)neutrino survival probabilities for Ar to H and shown in Figure \ref{fig8}. The upper panel of Figure \ref{fig8} is for $\nu$ and lower is for $\bar{\nu}$. From this figure, one notes that the ratios from both GENIE and NuWro have a lot of fluctuations at lower energy. We also note two sharp peaks, one around 0.5 GeV and another around 2.7 GeV. 
	The ratio of P(Ar)/P(H) calculated as a function of true neutrino energy (shown by blue stars) shows less fluctuations as compared to the ratio calculated with reconstructed neutrino energy (shown by red stars). Similar analysis is also performed with the older version (v2.12.10) of GENIE, and it is observed that results with the latest version (v3.0.6) of GENIE are better than the older version (v2.12.10) of GENIE.

	\section{Conclusions}
	
	\label{sec5}
	We should have a good understanding of the hadronic physics of neutrino-nucleus interactions since it is critical to understand nuclear effects. The information about the energy dependence of all exclusive cross-sections and nuclear effects is important for the construction of any nuclear model. Any ignorance of theoretical uncertainties in the nuclear models while simulating results cost inaccuracy. We use different event generators, which are built upon different nuclear models to test results from the neutrino oscillation experiments.
	
	Figure \ref{fig8} shows that the nuclear models of the two most extensively used neutrino event generators, GENIE and NuWro, have non-negligible systematic inaccuracies, especially in the energy range of 1-3 GeV. To augment our knowledge of the poorly understood nuclear effects, a study of high-statistics $\nu$-H scattering data is required. Through our work, we may conclude that  for long-baseline neutrino experiments like DUNE  a multi-target approach might be useful to reduce nuclear effects.
	
	\section{Acknowledgement.}
	One of the authors, Miss Ritu Devi offers most sincere gratitude to the Council of Scientific and Industrial Research (CSIR), Government of India, for the financial support in the form of Senior Research Fellowship, file no. 09/100(0205)/2018-EMR-I.

\end{document}